\documentclass[aps,pra,10pt,a4paper,twocolumn,footinbib, superscriptaddress]{revtex4-2}
\usepackage{graphicx}
\usepackage{amsmath}
\usepackage{amsfonts}
\usepackage{amssymb}
\usepackage{soul}
\usepackage{dsfont}
\usepackage{hyperref}
\usepackage{amstext}
\usepackage[caption=false]{subfig}
\usepackage{epstopdf} 
\usepackage{mathtools} 
\usepackage{mnsymbol} 
\usepackage{balance}
\hypersetup{colorlinks=true, citecolor=blue, urlcolor=blue, linkcolor=blue}
\usepackage{color,soul}

\usepackage{physics}
\usepackage{braket}

\newcommand{\RN}[1]{\textup{\uppercase\expandafter{\romannumeral#1}}}

\renewcommand{\rm}{\mathrm}
\newcommand{\kket}[1]{\left|#1\right\rrangle}
\newcommand{\bbra}[1]{\left\llangle#1 \right|}
\newcommand{\bbraket}[1]{\left\llangle #1 \right\rrangle}

\date{\today}

\begin{document}

\title{Master-equation treatment of nonlinear optomechanical systems with optical loss}

\author{Sofia Qvarfort }
\email{ sofiaqvarfort@gmail.com}
\affiliation{QOLS, Blackett Laboratory, Imperial College London,  SW7 2AZ London, United Kingdom}
\affiliation{Department of Physics and Astronomy, University College London, Gower Street, WC1E 6BT London, United Kingdom}
\author{Michael R. Vanner}
\affiliation{QOLS, Blackett Laboratory, Imperial College London,  SW7 2AZ London, United Kingdom}
\author{P. F. Barker}
\affiliation{Physics and Astronomy, University College London, Gower Street, WC1E 6BT London, United Kingdom}
\author{David Edward Bruschi}
\email{david.edward.bruschi@posteo.net}
\affiliation{Department of Theoretical Physics, Universit\"at des Saarlandes,  66123 Saarbr\"ucken, Germany}

\begin{abstract}
Open-system dynamics play a key role in the  experimental and theoretical study of cavity optomechanical systems. In many cases, the quantum Langevin equations have enabled excellent models for optical decoherence, yet a master-equation approach to the fully nonlinear optomechanical Hamiltonian has thus far proven more elusive. To address this outstanding question  and broaden the mathematical tool set available, we derive a solution to the Lindblad master equation that models optical decoherence for a system evolving with the nonlinear optomechanical Hamiltonian. The method combines a Lie-algebra solution to the unitary dynamics with a vectorization of the Lindblad equation, and we demonstrate its  applicability by considering the preparation of optical cat states via the optomechanical nonlinearity in the presence of optical loss. Our  results provide a direct way of analytically assessing the impact of optical decoherence on the optomechanical intracavity state. 
\end{abstract}

\maketitle

\section{Introduction} 

Recent years have seen significant interest in the study of theoretical and experimental aspects of optomechanical systems~\cite{aspelmeyer2014cavity}. In particular,  the reported achievements of ground-state cooling~\cite{chan2011laser,teufel2011sideband,delic2020cooling} as well as the entanglement of macroscopic systems~\cite{lee2011entangling,riedinger2018remote,ockeloen2018stabilized, thomas2020entanglement}, have significantly improved the prospects for using optomechanical systems as sensors~\cite{mason2019continuous,rademacher2020quantum, yu2020quantum} and for tests of fundamental physics~\cite{bose1999scheme,marshall2003towards, kleckner2008creating,Derakhshani2016,bose2017spin,marletto2017gravitationally}.

The central feature of optomechanical systems is the radiation-pressure interaction between light and matter, which allows for exquisite experimental readout and control. In most cavity-based experiments, the radiation pressure from the input laser couples the photon number to the center-of-mass motion of the mechanical element. This interaction is fundamentally \textit{nonlinear}~\cite{law1995interaction},~i.e., the interaction Hamiltonian is a product of three field operators and the resulting equations of motion for the optical and mechanical modes cannot be written as a linear system of equations.  

The dynamics of the nonlinear optomechanical Hamiltonian with a constant light--matter coupling was first solved by two pioneering theoretical studies~\cite{mancini1997ponderomotive, bose1997preparation}. The solutions inspired numerous proposals for tests of fundamental physics~\cite{bose1999scheme,marshall2003towards, kleckner2008creating}, sensing schemes~\cite{qvarfort2018gravimetry, armata2017quantum, schneiter2020optimal, qvarfort2021optimal}, and studies of the generation of non-Gaussian states~\cite{qvarfort2019enhanced, qvarfort2020time}. In many cases, however, the nonlinear optomechanical Hamiltonian is linearized around a strong coherent input state~\cite{aspelmeyer2014cavity}, which sacrifices the nonlinearity  and the ability to generate non-Gaussian states for a more tractable mathematical treatment~\cite{serafini2017quantum}. Indeed, most experiments to date are well-described by the linearized optomechanical Hamiltonian~\cite{aspelmeyer2014cavity}, but as an increasing number of theoretical~\cite{ludwig2008optomechanical, nunnenkamp2011single,rabl2011photon,vanner2011selective} and experimental works~\cite{brawley2016nonlinear,leijssen2017nonlinear} enable the study and observations of nonlinear phenomena, it becomes imperative to develop theoretical tools that accurately describe experiments in the nonlinear regime. 

An outstanding challenge involves a general and analytical treatment of optical decoherence in a nonlinear optomechanical system. Typically, open dynamics are modeled by solving either a master equation or the quantum Langevin equation~\cite{gardiner2004quantum}. Since the latter can be integrated into the input-output theory framework,  it has long been the main focus of the community. 
In contrast, modeling optical decay through a master equation has been generally challenging because the optical dissipation terms do not commute with the optomechanical interaction term. A perturbative solution for slowly decaying systems was taken as a first step by~\citet{mancini1997ponderomotive}. Mechanical loss, on the other hand, has been exactly modeled in terms of the Lindblad equation for phonon dissipation~\cite{bose1997preparation} and Brownian motion~\cite{bassi2005towards}. In addition, a treatment of both optical and mechanical losses through a damping-basis approach~\cite{briegel1993quantum} has also been put forward~\cite{torres2019optomechanical}. 

\begin{figure}[b!]
\includegraphics[width =0.3\textwidth, trim = 0mm 0mm 0mm 0mm]{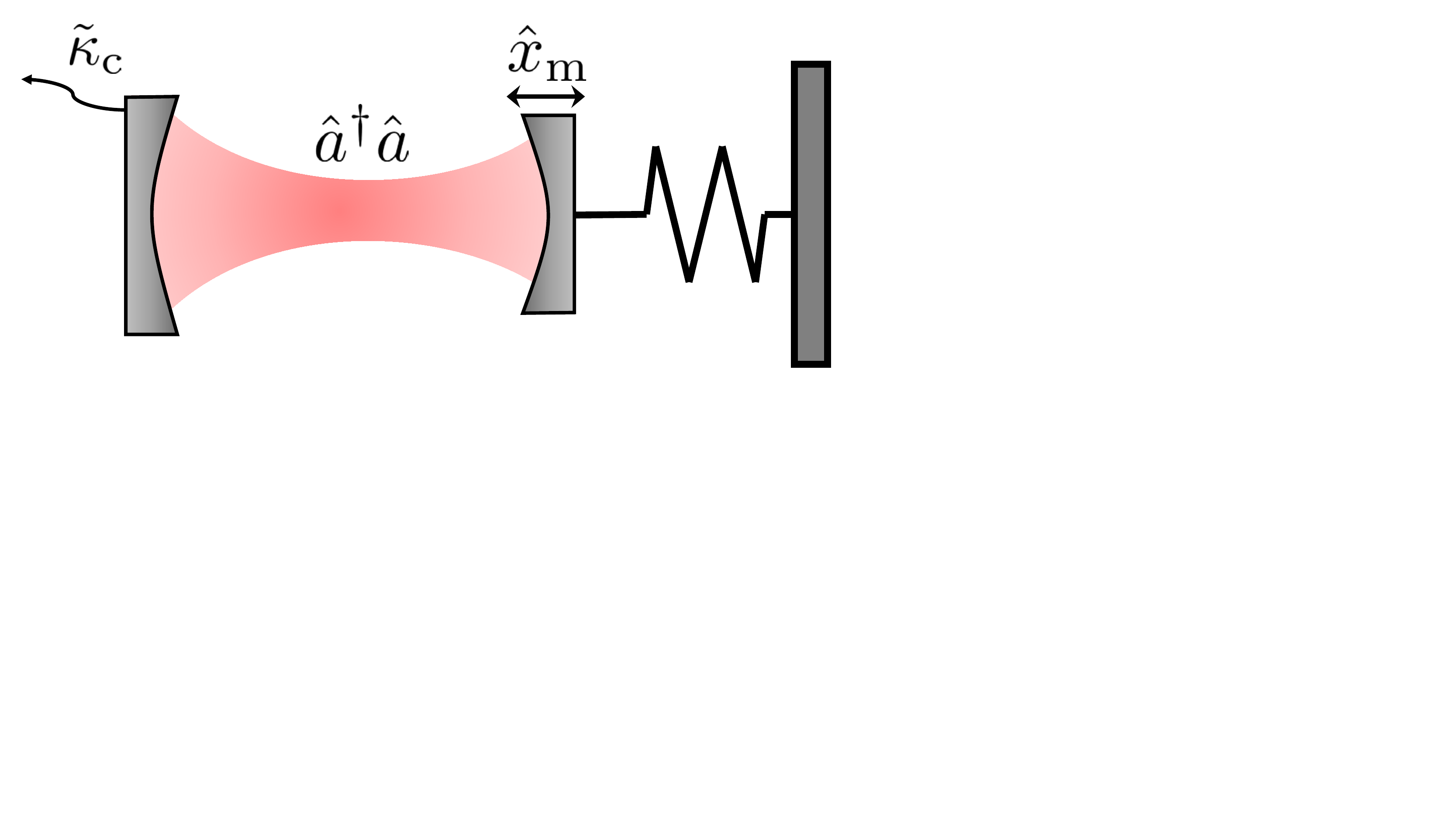}
\caption{Optomechanical setup where the optical mode $\hat a$ is coupled to the mechanical position $\hat x_{\rm{m}}$ via the interaction term $\hat a^\dag \hat a \, \hat x_{\rm{m}}$. Imperfections cause the photons to leak from the cavity at a rate $\kappa_{\rm{c}}$, which we represent as a rescaled number with respect to the mechanical frequency $\omega_{\rm{m}}$ as $\tilde{\kappa}_{\rm{c}} = \kappa_{\rm{c}}/\omega_{\rm{m}}$.}
\label{fig:cavity}
\end{figure}

In this work  we derive an expression for the nonunitary evolution of a nonlinear optomechanical system by combining a previously established Lie-algebra solution~\cite{wei1963lie} for the unitary dynamics~\cite{bruschi2018mechano,qvarfort2019enhanced} with a vectorization of the Lindblad equation. We also make use of the fact that the nonunitary evolution can be partitioned into separate products in a manner similar to that by which the interaction picture is utilized. To demonstrate how our solution to the Lindblad equation may be applied, we consider the preparation of optical cat states via the nonlinear optomechanical interaction in the presence of optical loss. Our results allow us to bound the optical decay rate given a desired fidelity with which we wish to prepare the states. 

The work is structured as follows. In Sec.~\ref{sec:unitary:dynamics}, we review the known unitary solutions for a nonlinear optomechanical system. Following that, in Sec.~\ref{sec:tools} we introduce the Lindblad equation along with the two methods we use for solving it: vectorization and partitioning the time evolution. We proceed to apply these methods  in Sec.~\ref{sec:optical:decoherence:nonlinear:optomechanics} to a nonlinear optomechanical system with optical decoherence and consider the three above examples in Sec.~\ref{sec:examples}. We conclude our work with a summary and outlook in Sec.~\ref{sec:conclusions}.

\section{Unitary dynamics of the nonlinear optomechanical Hamiltonian} \label{sec:unitary:dynamics}
We begin by considering a single mode of an optical field that is nonlinearly coupled to the center-of-mass mode of a mechanical element (see Fig.~\ref{fig:cavity}). The full Hamiltonian for the cavity mode and mechanical mode reads
\begin{align} \label{eq:Hamiltonian}
\hat H(t) &= \hbar \, \omega_{\rm{c}} \, \hat a ^\dag \hat a + \hbar \, \omega_{\rm{m}} \, \hat b^\dag \hat b - \hbar  \,  g(t) \,  \hat a^\dag \hat a\, \bigl( \hat b^\dag + \hat b \bigr), 
\end{align}
where $\omega_{\rm{c}}$ and $\omega_{\rm{m}}$ are the oscillation frequencies of the optical and mechanical modes respectively, and $g(t)$ denotes the (possibly time-dependent) light--matter coupling strength. The modes are defined by the annihilation and creation operators $\hat a, \hat a^\dag$ and  $\hat b, \hat b^\dag$, which satisfy the canonical commutator relations $[\hat a , \hat a^\dag ] = [\hat b, \hat b^\dag] = 1$.

For simplicity of notation, we proceed to rescale all frequencies by $\omega_{\rm{m}}$, which is equivalent to defining a dimensionless time parameter $\tau = t\,\omega_{\rm{m}}$. With this choice of notation, the optomechanical coupling  can be written as $\tilde{g}(\tau) = g(\tau/\omega_{\mathrm{m}})/\omega_{\rm{m}}$. We redefine the Hamiltonian $\hat H(t) \rightarrow \hat H(\tau)$ in these new dimensionless units as
\begin{align} \label{eq:Hamiltonian:rescaled}
\hat{H}(\tau) &= \hbar \, \frac{\omega_{\rm{c}}}{\omega_{\mathrm{m}}} \, \hat a ^\dag \hat a + \hbar \,  \hat b^\dag \hat b - \hbar  \,  \tilde{g}(\tau) \,  \hat a^\dag \hat a\, \bigl( \hat b^\dag + \hat b \bigr).
\end{align}
The time evolution operator that corresponds to~\eqref{eq:Hamiltonian:rescaled} is given by 
\begin{align} \label{eq:time:evolution}
\hat U(\tau) = \overleftarrow{\mathcal{T}} \mathrm{exp} \left[ - \frac{i}{\hbar } \int^\tau_0 \mathrm{d}\tau' \, \hat{H}(\tau') \right], 
\end{align}
where $\overleftarrow{\mathcal{T}}$ denotes the time ordering of the exponential.

A solution of~\eqref{eq:time:evolution} for a constant optomechanical coupling was derived by~\citet{bose1997preparation} and~\citet{mancini1997ponderomotive}. When the optomechanical coupling is time dependent, however, the solutions become more complex. It has been previously shown that a Lie-algebra method can be used to obtain solutions for general time dependence~\cite{bruschi2018mechano, qvarfort2019enhanced}. Here we summarize the results. 

By identifying a set of operators that is closed under commutation, the time-evolution operator $\hat U(\tau)$ in~\eqref{eq:time:evolution} can be written as
\begin{align} \label{eq:full:time:evolution}
\hat U(\tau)&= e^{- i \hat N_b  \tau} e^{-i F_{a}\,\hat{N}^2_a} e^{-i F_{+} \hat{N}_a\,\hat{B}_+}
 e^{-i F_{_-}\,\hat{N}_a\,\hat{B}_-} ,
\end{align}
where we have transformed into a frame that rotates with the free optical evolution $\mathrm{exp}\bigl[ -  i \hat a^\dag \hat a \, \tau \, \omega_{\mathrm{c}}/\omega_{\mathrm{m}} \bigr]$, and we have defined the following Hermitian operators: $
 \hat{N}_a := \hat a^\dagger \hat a$, $
	\hat{N}_b := \hat b^\dagger \hat b$, $\hat{B}_+ :=  \hat b^\dagger +\hat b$,  and $
	\hat{B}_- := i\,(\hat b^\dagger -\hat b)$. 
The $F$ coefficients in~\eqref{eq:full:time:evolution} are functions of time $\tau$ and are given by the following integrals~\cite{bruschi2018mechano, qvarfort2019enhanced}:
\begin{align} \label{eq:definition:of:F:coefficients}
&F_{a} =2   \int^\tau_0 \mathrm{d}\tau' \, \tilde{g}(\tau') \sin(\tau') \, \int^{\tau'}_0 \mathrm{d}\tau'' \tilde{g}(\tau'') \, \cos(\tau''),  \nonumber \\
&F_{+} = - \int^\tau_0 \mathrm{d}\tau' \tilde{g}(\tau')  \, \cos(\tau'), \,  \nonumber \\
&F_{-} = \int^\tau_0 \mathrm{d} \tau' \tilde{g}(\tau') \, \sin(\tau').
\end{align}
For a constant optomechanical coupling $\tilde{g}(\tau) \equiv \tilde{g}_0 = g_0/\omega_{\rm{m}}$,  the integrals in~\eqref{eq:definition:of:F:coefficients} evaluate to 
\begin{align} \label{eq:F:coefficients}
&\quad\quad\quad\quad F_a = \frac{1}{2}\tilde{g}_0^2 \left[  \sin(2\tau) -  2\tau \right],  \\
& F_+ = - \tilde{g}_0 \, \sin(\tau), && F_- = \tilde{g}_0  \, \left[ \cos(\tau) - 1 \right],\nonumber
\end{align}
which is equivalent to the previously obtained solutions~\cite{bose1997preparation, mancini1997ponderomotive} up to the ordering of the terms in~\eqref{eq:full:time:evolution}.

\section{The Lindblad equation} \label{sec:tools}
The Lindblad equation describes Markovian noise-processes as an effective nonunitary contribution to the dynamics~\cite{gardiner2004quantum}. The most general form of the  quantum master equation in Gorini-Kossakowski-Sudarshan-Lindblad form for $N$ environmental modes reads~\cite{lindblad1976generators, gorini1976completely}
\begin{equation} \label{eq:Lindblad:general}
\dot{\hat{\varrho}} = - i [\hat H, \hat \varrho] + \sum_{n,m=1}^{N^2-1} h_{nm} \left(  \hat L_n  \, \hat \varrho \, \hat L_m^\dagger - \frac{1}{2} \{ \hat L_m^\dagger \hat L_n , \hat \varrho \} \right), 
\end{equation}
where $\hat \varrho$ is the density matrix of a quantum state, $\hat H$ is the Hamiltonian operator,  $\hat L_n$ is a non-Hermitian Lindblad operator, and where  $\{\cdot, \cdot \}$ denotes the anti-commutator.

To obtain a solution to~\eqref{eq:Lindblad:general}, we make use of two methods:  vectorization and a factorization of the evolution operator akin to moving to the interaction picture. The combination of these methods allows us to write down a solution based on the previously obtained Lie-algebra solution for the unitary dynamics. We outline both methods in the following sections. 

\subsection{Introduction to vectorization}

Here we introduce the vectorization procedure for linear operators that act on the Hilbert space and show how the vectorized Lindblad equation is derived. We also refer the reader to the excellent introduction to vectorization in~\cite{d2000bell} and in the Supplemental Material of~\cite{alipour2014quantum}, the notation of which we follow closely. 

We start by considering a generic operator $\hat A$ that acts on the Hilbert space $\mathcal{H}$. Given an orthonormal basis $\{\ket{i}\}$ in  $\mathcal{H}$, the operator $\hat A$ can be written as
\begin{equation}
\hat A = \sum_{ij} \bra{i} \hat{A} \ket{j} \ketbra{i}{j}. 
\end{equation}
We then assign a vector to this operator by flipping one of the bras into a ket:
\begin{equation}
\kket{A} = \sum_{ij} \bra{i} \hat{A} \ket{j} \ket{i}\ket{j}.
\end{equation}
That is, every row in the matrix $\hat{A}$ defined through its elements $A_{ij}:=\bra{i} \hat{A} \ket{j}$ becomes stacked in the vector $\kket{A}$. We note that this makes the vectorization basis dependent. 

To vectorize the Lindblad equation, we need  the  relation (see~\cite{alipour2014quantum} for the derivation)
\begin{equation}\label{app:eq:key:vector:identity}
\kket{ABC} =  ( \hat A \otimes \hat C^{\rm{T}}) \kket{B},
\end{equation}
which demonstrates how a vectorized product of operators can be considered. We will later replace $\hat B$ by the density matrix in the Lindblad equation (see Sec.~\ref{sec:optical:decoherence:nonlinear:optomechanics}). 
Finally, we note that expectation value for the general operator $\hat A$ and the state $\hat \varrho$ is given in the vectorized language as
\begin{equation} \label{eq:overlap}
\braket{\hat A} = \mathrm{Tr}\bigl[ \hat A \, \hat \varrho \bigr] = \bbraket{A^\dag | \varrho}.
\end{equation}
This will allow us to compute various quantities of interest once we have solved the dynamics. 

\subsection{Vectorizing the Lindblad equation}
As noted in the preceding section, vectorization transforms matrices into vectors. Crucially, it also allows us to transform super operators into matrices. In fact, this method has been used to great effect in previous efforts to model nonunitary dynamics (see, e.g.,~\cite{alipour2014quantum, teuber2020solving, buvca2020bethe}).  

In this work we denote the vectorized density matrix $\hat \varrho $ by $\kket{\varrho}$ and the free state evolution is subsequently written as 
\begin{equation}
\hat U(t) \, \hat \varrho_0\, \hat U^\dag (t) \rightarrow \hat U(t) \otimes \hat U^*(t) \kket{\varrho_0}.
\end{equation}
Note that we here take the complex conjugate rather than the full conjugate transpose of $\hat U(t)$, as mandated by the vectorization mapping that we chose. Throughout this work, we use the tensor product to differentiate between the left-hand and right-hand multiplication of $\hat U(t)$ throughout, rather than showing the structure of the Hilbert space in terms of the optical and mechanical modes.

\begin{table}
\caption{\label{tab:vectorized} Vectorized analogs of terms in the Lindblad equation~\eqref{eq:Lindblad:general}.}
\begin{ruledtabular}
\begin{tabular}{cc}
Operator product & Vectorized analog \\\hline
$\hat \varrho \, \hat H $ &  $(\mathds{1}\otimes \hat H^{\mathrm{T}}) |\varrho \rrangle $ \\
 $\hat H \, \hat \varrho$ & $(\hat H\otimes \mathds{1}) |\varrho \rrangle$\\
$\hat L_n \,  \hat  \varrho \, \hat L^\dagger_m$ & $(\hat L_n \otimes \hat L_m^{\dagger \mathrm{T}} ) |\varrho \rrangle$ \\
$\hat L_m^\dagger \hat L_n \, \hat \varrho $ &  $(\hat L_m^\dagger \hat L_n \otimes \mathds{1} ) |\varrho \rrangle $ \\
$\hat \varrho \, \hat L^\dagger_m \hat L_n$ & $  (\mathds{1} \otimes (\hat L_m^\dagger \hat L_n)^{\mathrm{T}} ) |\varrho \rrangle$
\end{tabular}
\end{ruledtabular}
\end{table}

To apply the vectorization to the Lindblad equation, 
we use the identity~\eqref{app:eq:key:vector:identity} on all terms of the Lindblad equation~\eqref{eq:Lindblad:general}.   The terms and their vectorized analogs can be found  in Table~\ref{tab:vectorized}. Where only two operators were multiplied, we inserted the identity to ensure that we obtain products of three operators. As a result,~\eqref{eq:Lindblad:general} can be written in the vectorized language as
\begin{equation} \label{eq:vectorized:Lindblad}
\frac{\mathrm{d}}{\mathrm{d} t}|\varrho \rrangle  = \hat{\mathcal{L}}(t)  \kket{\varrho}.
\end{equation}
Here we write $\hat{\mathcal{L}}(t)$ as:
\begin{equation} \label{eq:vectorized:Lindblad:super:operator}
\hat{\mathcal{L}} =  \hat{\mathcal{L}}_H + \hat{\mathcal{L}}_L,
\end{equation} 
where (according to the terms listed in Table~\ref{tab:vectorized}) $\hat{\mathcal{L}}_H$ is the unitary (Hamiltonian) contribution given by $\hat{\mathcal{L}}_H := - i \bigl( \hat H(t) \otimes \mathds{1} -  \mathds{1}\otimes \hat H^{\rm{T}}(t) \bigr) $ and  $\hat{\mathcal{L}}_L$ contains the nonunitary part
\begin{align} \label{eq:LH:LL:definitions}
\hat{\mathcal{L}}_L &:=   \sum_{n,m = 1}^{N^2-1} \frac{h_{nm}}{2} \left[ 2 \hat L_n\otimes \hat L_m^{\dagger \rm{T}}- \hat L_m^\dagger \hat L_n \otimes \mathds{1} + \mathds{1}\otimes  (\hat L_m^\dagger \hat L_n)^{\rm{T}} \right] . 
\end{align}
These expressions might appear nonintuitive at first because of the notation used for the vectorization. The vectorization essentially splits the system into two modes (here explicitly indicated by use of the tensor product), one 'right-handed' and one 'left-handed' mode, which act on separate parts of the vectorized density matrix. We also notice the appearance of transposed operators in~\eqref{eq:LH:LL:definitions}, which follow from our choice of the vectorization mapping. However, we may simplify the expression by adopting a real basis, such as the Fock basis, where $\hat L$ and $\hat L^\dagger$ have exclusively real entries. This means that the transposition operation is equivalent to taking the Hermitian conjugate, which, for example, allows us to write $\hat L^{\rm{T}}_i= \hat L_i^\dag$.  This will greatly simplify our calculations, but may have consequences for the case where we wish to explicitly compute quantities using a complex basis. We do not, however, encounter those cases in this work. 

The formal solution to the Lindblad equation~\eqref{eq:vectorized:Lindblad} in the vectorized language reads
\begin{equation} \label{eq:Lindblad:formal:solution}
\kket{\varrho(t)} = \hat{\mathcal{S}}(t) \kket{\varrho_0},
\end{equation}
where $\kket{\varrho_0}$ is the vectorized form of the initial state $\hat \varrho_0$ and $\hat{\mathcal{S}}(t)$ is the time-ordered exponential of $\hat{\mathcal{L}}(t)$:
\begin{equation} \label{eq:formal:solution:S}
\hat{\mathcal{S}}(t) = \overleftarrow{\mathcal{T}} \mathrm{exp} \left[ \int^t_0 \mathrm{d} t' \, \hat{\mathcal{L}}(t') \right]. 
\end{equation}
This is a key expression that captures both the unitary and the nonunitary evolution. In the next section, we proceed to show how $\hat{\mathcal{S}}(t)$ may be further simplified.

\subsection{Partitioning the time-evolution} \label{sec:partitioned:time:evolution}

The second method that we will use to solve the Lindblad equation relies on the fact that any time-evolution operator $\hat U(t)$ [or the nonunitary evolution operator $\hat{\mathcal{S}}(t)$, as will become evident] can be partitioned into products that arise from the different Hamiltonian terms. Once partitioned, each contribution can then be evaluated using a suitable method. For example, the time evolution that arises from a quadratic Hamiltonian can be treated using phase-space methods~\cite{serafini2017quantum}, while a cubic or higher Hamiltonian term can in some cases be treated with a Lie-algebra method~\cite{wei1963lie}, as we do here. 

Formally, we consider the time-evolution operator $\hat U(t)$ generated by the Hamiltonian $\hat H(t) = \hat H_A(t) + \hat H_B(t)$, where the partition of $\hat H_A(t) $ and $\hat H_B(t)$ is arbitrary. We may then consider a frame that rotates with $\hat U_A(t)$, which is defined in the standard way as
\begin{equation}
\hat U_A(t) = \overleftarrow{\mathcal{T}} \mathrm{exp}\left[ - \frac{i}{\hbar } \int^t_0 \mathrm{d}t' \, \hat H_A(t') \right].
\end{equation}
It is then possible to write $\hat U(t)$ as the following product $
\hat U(t) = \hat U_A(t) \, \hat U_B(t)$,
where $\hat U_B(t) $ is given by
\begin{equation} \label{eq:UB:evolution}
\hat U_B(t) = \overleftarrow{\mathcal{T}}\mathrm{exp}\left[ - \frac{i}{\hbar } \int^t_0 \mathrm{d}t' \, \hat U_A^\dag (t') \, \hat H_B (t')\, \hat U_A(t') \right]. 
\end{equation}
See Appendix~\ref{app:interaction:picture} for a detailed derivation, which follows the standard treatment of the interaction picture. 

We now seek to generalize these notions to nonunitary dynamics. Consider  $\hat{\mathcal{L}}(t) = \hat{\mathcal{L}}_A(t) + \hat{\mathcal{L}}_B(t)$, where again $\hat{\mathcal{L}}_A(t)$ and $\hat{\mathcal{L}}_B(t)$ are completely arbitrary. The formal solution for the evolution with $\hat{\mathcal{L}}_A(t)$ is given from~\eqref{eq:formal:solution:S}  and reads
\begin{equation}
\hat{\mathcal{S}}_A(t) = \overleftarrow{\mathcal{T}} \mathrm{exp}\left[ \int^t_0 \mathrm{d}t'\, \hat{\mathcal{L}}_A(t') \right], 
\end{equation}
and by considering a transformation similar to the interaction picture for unitary dynamics, we write $\hat{\mathcal{S}}(t) = \hat{\mathcal{S}}_A(t) \, \hat{\mathcal{S}}_B(t)$, where now
\begin{equation} \label{eq:simplified:S}
\hat{\mathcal{S}}_B(t) = \overleftarrow{\mathcal{T}} \mathrm{exp} \left[ \int^t_0 \mathrm{d}t' \, \hat{\mathcal{S}}^{-1}_A (t') \, \hat{\mathcal{L}}_B (t') \, \hat{\mathcal{S}}_A (t') \right].
\end{equation}
In some cases, partitioning $\hat{\mathcal{S}}(t)$  in this way simplifies the problem at hand. We provide a formal proof of the fact that the partitioning holds for nonunitary dynamics in Appendix~\ref{app:interaction:picture}. 

We are now ready to consider the Lindblad master equation for optical decoherence in a nonlinear optomechanical system.

\section{Optical decoherence in a nonlinear optomechanical system} \label{sec:optical:decoherence:nonlinear:optomechanics}

The main loss mechanisms in an optical cavity are comprised of intrinsic losses, such as scattering and absorption, and of extrinsic losses, such as an imperfect mirror reflectivity or losses from the output coupling~\cite{aspelmeyer2014cavity}. The latter can generally be controlled in experiments, while the former are unavoidable. We call the total decay rate $\kappa_{\mathrm{c}}$, which gives rise to dissipation in the energy basis, which in turn leads to decoherence of the off-diagonal elements in the density matrix. 

Our goal is to solve the Lindblad equation for optical decoherence in a nonlinear optomechanical system. 
Concretely, we wish to derive an expression for $\hat{\mathcal{S}}(\tau)$ [shown in~\eqref{eq:Lindblad:formal:solution}] that can be used to evaluate quantities of interest.  To do so, we start from the vectorized Lindbladian~\eqref{eq:vectorized:Lindblad:super:operator} for a single optical mode
\begin{align} \label{eq:Lindbladian}
\hat{\mathcal{L}}(\tau) &= - i\left[ \hat H(\tau) \otimes \mathds{1} -  \mathds{1}\otimes \hat H (\tau) \right]   + \hat L \otimes \hat L \nonumber \\
&\qquad - \frac{1}{2} \left( \hat L^\dagger \hat L \otimes \mathds{1} + \mathds{1} \otimes \hat L^\dagger \hat L \right), 
\end{align}
where $\hat H(\tau)$ is the optomechanical Hamiltonian rescaled by $\omega_{\mathrm{m}}$ shown in~\eqref{eq:Hamiltonian:rescaled}. To model optical dissipation, we let the Lindblad operator be $\hat L = \sqrt{\tilde{\kappa}_{\rm{c}}} \, \hat a$, where $\tilde{\kappa}_{\rm{c}} = \kappa_{\rm{c}}/\omega_{\rm{m}}$ is the rescaled optical damping rate. 

We proceed by partitioning~\eqref{eq:Lindbladian} into the following unitary and nonunitary parts:
\begin{align}
\hat{\mathcal{L}}_H &= i \, \mathds{1}\otimes \hat H(\tau) - i \, \hat H(\tau) \otimes \mathds{1},  \nonumber \\
\hat{\mathcal{L}}_L &= \frac{\tilde{\kappa}_{\rm{c}}}{2} \,  \left( 2 \, \hat a \otimes \hat a -   \hat N_a \otimes \mathds{1} + \mathds{1} \otimes  \hat N_a \right).
\end{align}
This partition  allows us to write the full solution to the Lindblad equation~\eqref{eq:formal:solution:S}  as $\hat{\mathcal{S}}(\tau) = \hat{\mathcal{S}}_H(\tau) \,  \hat{\mathcal{S}}_L(\tau)$ (see Sec.~\ref{sec:partitioned:time:evolution}),
where  
\begin{align} \label{eq:def:of:SH:SL}
\hat{\mathcal{S}}_H  &:= \overleftarrow{\mathcal{T}} \mathrm{exp} \left[ \int^\tau_0 \mathrm{d}\tau' \, \hat{\mathcal{L}}_H \right],  \nonumber \\ 
\hat{\mathcal{S}}_L  &:= \overleftarrow{\mathcal{T}} \mathrm{exp} \left[ \int^\tau_0 \mathrm{d} \tau' \,  \hat{\mathcal{S}}_H^{-1} \, \hat{\mathcal{L}}_L(\tau') \,  \hat{\mathcal{S}}_H  \right].
\end{align}
Note that $\hat{\mathcal{S}}_H(\tau)$ encodes the unitary evolution, since
\begin{align} \label{eq:unitary:part}
\hat{\mathcal{S}}_H(\tau) &= \overleftarrow{\mathcal{T}} \mathrm{exp} \left[i \int^\tau_0 \mathrm{d} \tau' \, \left[ \mathds{1} \otimes \hat H(\tau') - \hat H(\tau')\otimes \mathds{1} \right] \right]  \nonumber \\
&= \hat U(\tau) \otimes \hat U^*(\tau), 
\end{align}
where the solution of $\hat U(\tau)$ is shown in~\eqref{eq:full:time:evolution}. The additional complex conjugate arises from the choice of the  vectorization mapping. 

We then once again split $\hat{\mathcal{L}}_L$ into the following two components: $\hat{\mathcal{L}}_{\hat a, \hat a } = \tilde{\kappa}_{\rm{c}} \, \hat a \otimes \hat a$ and 
\begin{align}
\hat{\mathcal{L}}_{\hat N_a} = -\frac{1}{2} \tilde{\kappa}_{\rm{c}} \left( \mathds{1} \otimes\hat N_a +  \hat N_a \otimes \mathds{1}  \right). 
\end{align}
Then, using the fact that $\hat N_a$ commutes with the Hamiltonian~\eqref{eq:Hamiltonian}, we write the full solution as  $\hat{\mathcal{S}}(\tau)  = \hat{\mathcal{S}}_H \, \hat{\mathcal{S}}_{\hat N_a} \, \hat{\mathcal{S}}_{\hat a} $ where $\hat{\mathcal{S}}_H $ is defined in~\eqref{eq:def:of:SH:SL} and $\hat{\mathcal{S}}_{\hat N_a}$ and $\hat{\mathcal{S}}_{\hat a }$ are given by
\begin{align} \label{eq:Sa}
\hat{\mathcal{S}}_{\hat N_a} &= e^{- \tilde{\kappa}_{\rm{c}}  \tau \, \hat N_a /2} \otimes e^{- \tilde{\kappa}_{\rm{c}}  \tau \, \hat N_a/2},  \nonumber \\
\hat{\mathcal{S}}_{\hat a } &= \overleftarrow{\mathcal{T}} \mathrm{exp} \left[ \int^\tau_0 \mathrm{d}\tau' \, \hat{\mathcal{S}}_{\hat N_a}^{-1} \, \hat{\mathcal{S}}_H ^{-1} \, \hat{\mathcal{L}}_{\hat a, \hat a} \, \hat{\mathcal{S}}_H \, \hat{\mathcal{S}}_{\hat N_a} \right] \, .
\end{align}
To compute $\hat{\mathcal{S}}_{\hat a}(\tau)$, we must first examine the nontrivial term  $\hat{\mathcal{S}}^{-1}_H(\tau) \left( \hat a\otimes \hat a \right) \hat{\mathcal{S}}_H(\tau) $. Using~\eqref{eq:unitary:part}, we write  $
\hat{\mathcal{S}}^{-1}_H(\tau) \left( \hat a\otimes \hat a \right) \hat{\mathcal{S}}_H(\tau)  
= \hat U^\dag (\tau) \,  \hat a \, \hat U(\tau) \otimes  \hat U^{\dag*}(\tau) \, \hat a \, \hat U^* (\tau) $, where we again recall that we have disregarded the free optical evolution and that we used a basis where $\hat a$ has real entries, such that $\hat U^{\dag *}(\tau) \, \hat a \, \hat U^*(\tau) = \bigl[ \hat U^\dag (\tau) \, \hat a \, \hat U(\tau) \bigr]^*$. 
These terms are just the usual unitary Heisenberg evolution of $\hat a$, which is given by~\cite{qvarfort2019enhanced} 
\begin{align}\label{eq:evolution:of:a}
 \hat U^\dag (\tau) \, \hat a\, \hat U(\tau) &= e^{- i \, F_a }\, e^{- 2 \, i \, (F_a + F_+ F_-)\, \hat{N}_a} e^{- i F_+ \, \hat{B}_+ } \, e^{- i F_- \, \hat{B}_- } \, \hat{a}.
\end{align}
Then, since  $[\hat{\mathcal{S}}_{\hat N_a}(\tau), \hat U(\tau)]  =  0$, the term under the integral can be written
\begin{align} \label{app:eq:two:side:evolution}
&\left( e^{ -\tilde{\kappa}_{\rm{c}} \tau \hat N_a \otimes \mathds{1}/2} \, e^{ - \tilde{\kappa}_{\rm{c}}  \tau \mathds{1} \otimes \hat N_a /2 } \right)^{-1} \hat a\otimes \hat a \, e^{- \frac{1}{2} \tilde{\kappa}_{\rm{c}} \tau  \hat N_a \otimes \mathds{1}} \, e^{- \frac{1}{2}  \tilde{\kappa}_{\rm{c}} \tau   \mathds{1} \otimes \hat N_a } \nonumber \\
&= e^{-\tilde{\kappa}_{\rm{c}} \tau}  \, \hat a \otimes   \hat a ,
\end{align}
where we have used the relation $( \hat N_a)^n \, \hat a = \hat a \, ( \hat N_a - 1 )^n $, which in turn yields  $
e^{x \, \hat N_a} \, \hat a \, e^{-x \, \hat N_a}  = \, e^{-x} \, \hat a \, $.

Inserting the expressions~\eqref{eq:evolution:of:a} and~\eqref{app:eq:two:side:evolution} into $\hat{\mathcal{S}}_{\hat a}$~\eqref{eq:Sa}, we are able to write the full expression for $\hat{\mathcal{S}}(\tau)$ as 
\begin{widetext}
\begin{align} \label{eq:explicit:main:result}
\hat{\mathcal{S}}(\tau) &= \left( e^{- i \hat N_b\, \tau} \, e^{- i F_a \, \hat N_a^2} \, e^{- i F_+ \hat N_a \, \hat B_+} \, e^{- i F_- \hat N_a \, \hat B_-}e^{-  \tilde{\kappa}_{\rm{c}} \tau \hat N_a/2 }\right)   \otimes \left( e^{i \hat N_b\, \tau} \, e^{i F_a \, \hat N_a^2} \, e^{i F_+ \hat N_a \, \hat B_+} \, e^{i F_- \hat N_a \, \hat B_-}e^{- \tilde{\kappa}_{\rm{c}} \tau  \hat N_a/2 }\right) \nonumber \\
&\quad \times \overleftarrow{\mathcal{T}} \mathrm{exp} \left[\tilde{\kappa}_{\rm{c}} \int^\tau_0 \mathrm{d}\tau'  \, e^{- \tilde{\kappa}_{\rm{c}} \tau'} \,e^{-2\,i\,(F_a+F_+ F_- )\,\hat{N}_a}\,e^{-i\,F_+\,\hat{B}_+}\,e^{-i\,F_-\,\hat{B}_-}  \,  \hat a \otimes e^{2\,i\,(F_a+F_+ F_- )\,\hat{N}_a}\,e^{i\,F_+\,\hat{B}_+}\,e^{i\,F_-\,\hat{B}_-}  \, \hat a \,  \right], 
\end{align}
\end{widetext}
where the $F$ coefficients can be found in~\eqref{eq:definition:of:F:coefficients}, above which we have also listed the definitions of the operators.  We note that all $F$ coefficients inside the integral in~\eqref{eq:explicit:main:result} are functions of $\tau'$. 

To further simplify~\eqref{eq:explicit:main:result}, we write the operators under the integral that are acting on the mechanical subsystem as Weyl displacement operators
\begin{align} \label{eq:rewrite:displacement}
e^{- i \, F_+ \hat B_+ } e^{- i \, F_- \hat B_-} &=  \hat D(G(\tau)) \, e^{-  i \, F_+ F_-} , 
\end{align}
where we have defined $G(\tau) = F_- - i F_+$ and  the explicit form of the displacement operator is $\hat D(G(\tau)) \equiv e^{G(\tau) \hat b^\dag  - G^*(\tau) \hat b }$. 
Since the integral in~\eqref{eq:explicit:main:result} contains both $\hat D(G(\tau)) \, e^{-  i \, F_+ F_-}$ and its complex conjugate, we find that the phases cancel and that the final expression can be written in the  compact form
\begin{align}\label{eq:noisy:dynamics:general}
&\hat{\mathcal{S}}(\tau)  = \hat U(\tau) \, e^{-  \tilde{\kappa}_{\rm{c}}  \tau \,\hat N_a/2}  \otimes \hat U^*(\tau) \, e^{-  \tilde{\kappa}_{\rm{c}}  \tau \,\hat N_a /2 }  \nonumber \\
&\quad \times \overleftarrow{\mathcal{T}} \mathrm{exp} \biggl[\tilde{\kappa}_{\rm{c}} \int^\tau_0 \mathrm{d}\tau'  e^{- \tilde{\kappa}_{\rm{c}} \tau'}   \hat D(G(\tau')) \, e^{- 2  i  A(\tau')   \, \hat N_a}  \hat a \nonumber \\
&\qquad\qquad\qquad\quad \otimes \hat D(G^* (\tau')) \,e^{2  i  A(\tau') \,\hat N_a}  \,  \hat a  \biggr], 
\end{align}
where we have defined $A(\tau) = F_a + F_+ \, F_-$ and $\hat U(\tau)$ can be found  in~\eqref{eq:full:time:evolution}. 

Equation~\eqref{eq:noisy:dynamics:general} is the main result of this paper. It allows for optical dissipation to be considered for any rescaled coupling strength $\tilde{g}_0$ and any decay rate $\tilde{\kappa}_{\rm{c}}$. It generalizes a previous first-order perturbative solution for small $\tilde{\kappa}_{\mathrm{c}}$ derived by~\citet{mancini1997ponderomotive}. While~\eqref{eq:noisy:dynamics:general} cannot  be written in terms of a closed-form expression, \footnote{This is due the fact that the Lie algebra that generates the nonunitary evolution is infinite~\cite{wei1963lie}.  We can see this by commuting the terms in $\hat{\mathcal{L}}$, which leaves us with terms of the form proportional to $ \bigl(\hat b^\dag + \hat b \bigr)^N$, where $N \in \{1, \infty\}$.} we show in the following sections that it does in fact allow for certain quantities of the system to be computed.

\begin{figure*}
\centering
  \begin{minipage}{.45\textwidth}
        \centering
\subfloat[ \label{fig:homodyne}]{%
  \includegraphics[width=0.67\linewidth, trim = 7mm 0mm -7mm 0mm]{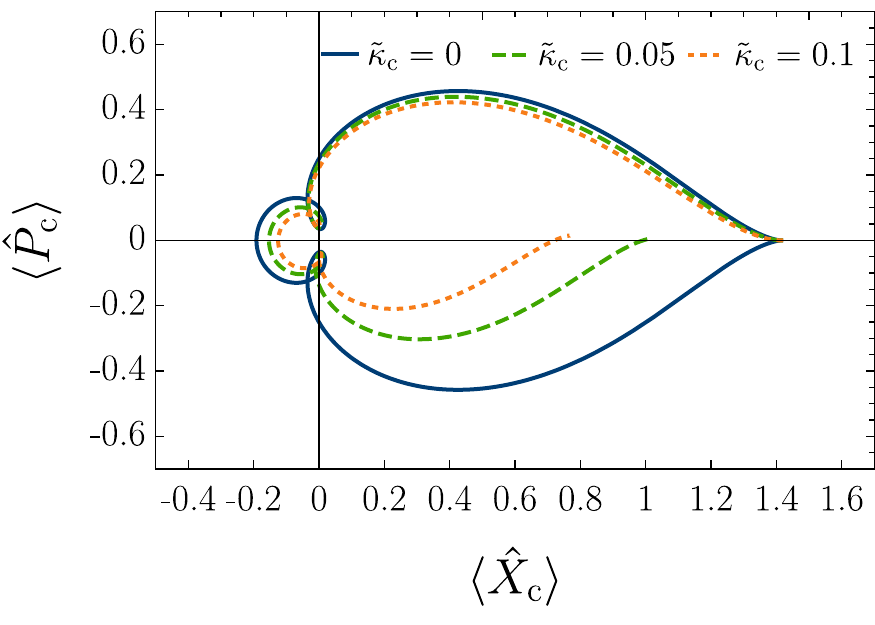}%
}  \\
\subfloat[ \label{fig:fidelity}]{%
  \includegraphics[width=0.67\linewidth, trim = 7mm 0mm -7mm 0mm]{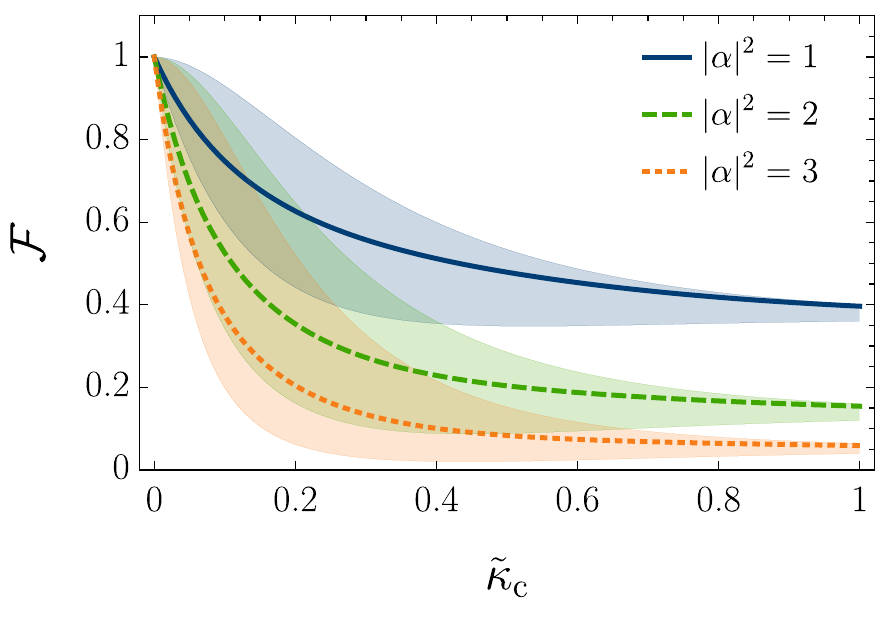}%
}
    \end{minipage}%
\hfill
  \begin{minipage}{.55\textwidth}
\subfloat[ \label{fig:Wigner}]{%
  \includegraphics[width=\linewidth, trim = 0mm -1mm 0mm 5mm]{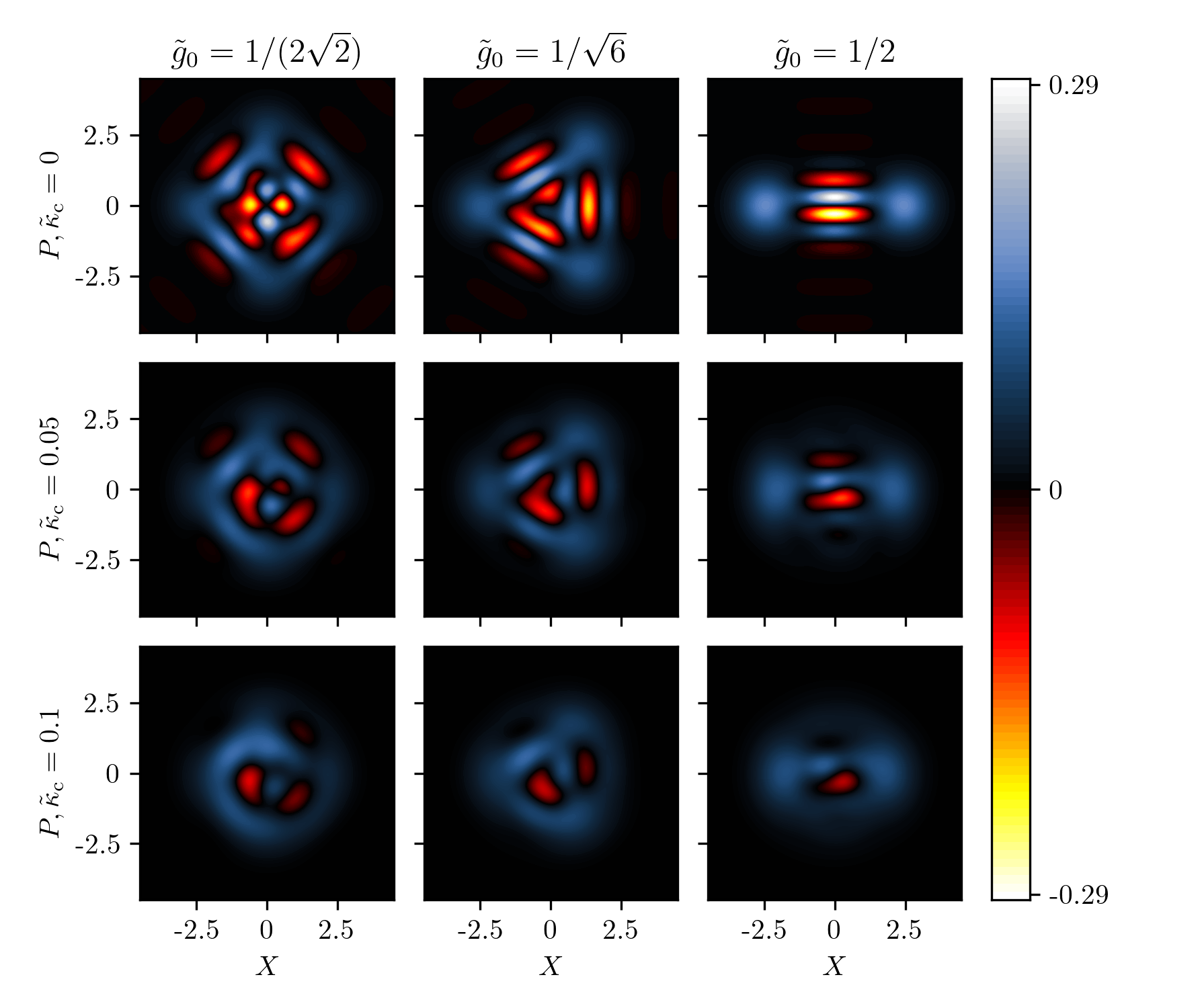}%
}
    \end{minipage}%
\caption{Impact of optical loss in a nonlinear optomechanical system.  (a) Parametric plot of the optical quadratures $\braket{\hat X_{\rm{c}}} = \sqrt{2} \, \mathrm{Re} [\braket{\hat a}]$ and $\braket{\hat P_{\rm{c}}} = \sqrt{2} \, \mathrm{Im}[\braket{\hat a}]$ as a function of time $\tau= \omega_{\mathrm{m}}t$ for $ \tilde{g}_0 = 1$. The optical state is a coherent state with $|\alpha| = 1$ and the mechanical mode is in the ground state. 
Time $\tau$ starts at the rightmost tip of the phase-space diagram and runs until $2\pi$. The expression for $\braket{\hat a}$ is given in~\eqref{eq:homodyne:signal}. For unitary dynamics (blue trajectory), the system returns to its initial state at $\tau = 2\pi$, while for nonunitary dynamics (green and orange lines), the quadratures decay and the system does not return to its original state. The larger $\tilde{\kappa}_{\mathrm{c}}$ is, the faster the state decays towards the vacuum expectation value.   (b) Fidelity $\mathcal{F}$ for generating a two-component optical cat state as a function of the decay rate $\tilde{\kappa}_{\rm{c}}$. The lines show the fidelity for three different coherent state parameters $\alpha$ given an optomechanical coupling of value $\tilde{g}_0 = \frac{1}{2}$. The shaded regions indicate the lower and upper bounds to $\mathcal{F}$ shown in~\eqref{eq:fidelity:bounds}.   (c) Grid of 3 $\times$ 3 numerically computed Wigner functions $W(X,P)$ of a noisy optical cat state with $|\alpha| = 3$. The coupling $\tilde{g}_0$  increases along the horizontal direction (from left to right) and the rescaled decay rate $\tilde{\kappa}_{\rm{c}}$ increases down along the vertical direction (from top to bottom). The negative values of $W(X,P)$, shown in red, indicate where the state is nonclassical. Even for $\tilde{\kappa}_{\rm{c}} \sim 0.05$ (middle row), the nonclassicality (red regions) rapidly decreases.}
\label{fig:}
\end{figure*}

\section{Examples} \label{sec:examples}

To demonstrate the utility of our method, we proceed to compute three quantities of interest: (i) the photon-number expectation value $\braket{\hat N_a}$, (ii) the intracavity quadratures, and (iii) the fidelity $\mathcal{F}$ for generating optical intracavity cat states in the presence of optical loss. 
In all three examples,  we work with the initially separable state of the mechanical and optical mode
\begin{equation} \label{eq:initial:state}
\ket{\Psi_0} = \ket{\alpha}_{\rm{c}}\otimes\ket{\beta}_{\rm{m}}, 
\end{equation}
where both $\ket{\alpha}_{\rm{c}}$ and $\ket{\beta}_{\rm{m}}$ are coherent states that satisfy the relations $\hat a\ket{\alpha}_{\rm{c}} = \alpha \ket{\alpha}_{\rm{c}}$ and $\hat b\ket{\beta}_{\rm{m}} = \beta \ket{\beta}_{\rm{m}}$. 

While it is commonly assumed that the representation of the optical state as a coherent state is an accurate one, the mechanical state is more often found in a thermal state, which is given by 
\begin{equation} \label{eq:thermal:state}
\hat \varrho_{\mathrm{th}} = \frac{1}{\pi \bar{n}} \int_{\mathbb{C}} \mathrm{d}^2 \beta \, e^{- |\beta|^2/\bar{n}} \ketbra{\beta}, 
\end{equation}
where $\beta \in \mathbb{C}$ and $\bar{n}$ is the average phonon number of the state. Often, results for~\eqref{eq:thermal:state} can be straighforwardly obtained by starting with the coherent state in~\eqref{eq:initial:state} and then integrating over $\beta$ with the appropriate weighting. We show below how this can be done for the intracavity optical quadratures of the state. In general, however, starting with an initial thermal state does not significantly further complicate the calculations because the vectorization has been chosen specifically to model mixed states.

\subsection{Photon-number}
For our first example, we  compute the expectation value of the photon-number operator $\hat N_a(\tau)$. Using the identity in~\eqref{eq:overlap}, we find  $\braket{\hat N_a} =  \llangle n | \hat{\mathcal{S}}(\tau)| \Psi_0 \rrangle$, where $\kket{n}$ is a vectorized Fock state (the eigenstate of $\hat N_a$) and  $\kket{\Psi_0}$ is the initial state shown in~\eqref{eq:initial:state}. 

The full calculations can be found in Appendix~\ref{app:photon:number}. The key step involves expanding the time-ordered exponential in~\eqref{eq:explicit:main:result} as a von Neumann series, and then acting on the various terms with the optical Fock states. We are then able to trace out the mechanical subsystem and contract the exponential again.  We are left with the relatively simple result
\begin{equation} \label{eq:photon:decay}
\braket{\hat N_a(\tau)} = |\alpha|^2 \, e^{-\tilde{\kappa}_{\rm{c}} \, \tau}, 
\end{equation}
where $|\alpha|^2$ is the initial number of photons in the cavity.  The photon number exponentially decays from the initial value $|\alpha|^2$ towards the vacuum state as $\tau \rightarrow \infty$ and corresponds exactly with numerical results. 
One might perhaps have expected the interaction between the optical and mechanical modes to influence $\braket{\hat N_a}$. However we note that $\hat N_a$ is a constant of the motion, which means that it commutes with the light-matter interaction term, and thus $\braket{\hat N_a(\tau)}$ decays just like a coherent state in a cavity would.

\subsection{intracavity optical quadratures}
The optical quadratures $\hat X_{\rm{c}} = \bigl( \hat a^\dag + \hat a \bigr) / \sqrt{2}$  and $\hat P_{\rm{c}} = i \bigl( \hat a^\dag - \hat a \bigr)/\sqrt{2}$ are the dimensionless first moments of the optical state. They are often measured in experiments using homodyne measurements and offer insights into the phase space trajectory of the systems. 

Our goal is to compute the expectation values $\braket{\hat X_{\rm{c}}(\tau)}$ and $\braket{\hat P_{\rm{c}}(\tau)}$. They are given in terms of the expectation value $\braket{\hat a(\tau)}$ as $\braket{\hat X_{\rm{c}}(\tau)} = \sqrt{2} \,  \mathrm{Re} \braket{\hat a(\tau)}$ and $\braket{\hat P_{\rm{c}}(\tau)} = \sqrt{2} \, \mathrm{Im} \braket{a(\tau)}$. Again using the identity~\eqref{eq:overlap}, we find that $\braket{\hat a}$ is given by $\braket{\hat a (\tau)} = \mathrm{Tr} \left[ \hat a  \, \hat \varrho( \tau)\right] = \bbraket{\hat a^\dag  | \varrho(\tau)} $. After again expanding the time-ordered exponential in~\eqref{eq:explicit:main:result} and effectively tracing out the mechanics (see Appendix~\ref{app:homodyne} for the full calculation), we find
\begin{align} \label{eq:homodyne:signal}
&\braket{\hat a(\tau)} =\alpha \, e^{|\alpha|^2 \left(e^{ - 2\, i \, A(\tau)} \, e^{- \tilde{\kappa}_{\rm{c}}  \tau}-1\right) } e^{-|G(\tau)|^2/2}\, \nonumber \\
&\quad \times  e^{- i\, A(\tau)}  \, e^{ -  \tilde{\kappa}_{\rm{c}} \tau/2} \, e^{G(\tau) \beta^* - G^*(\tau) \beta}  \, 	\nonumber \\
&\quad\times \mathrm{exp} \left[\tilde{\kappa}_{\rm{c}} \, |\alpha|^2 \int^\tau_0 \mathrm{d}\tau'  \, e^{- \tilde{\kappa}_{\rm{c}} \tau'} \,e^{-2\,i\,A(\tau')}  \, e^{i \, B(\tau', \tau) } \right], 
\end{align}
where we have defined $B(\tau', \tau) =  2\, \mathrm{Im}[G(\tau) G^*(\tau')]$. While the integral in~\eqref{eq:homodyne:signal} does not appear to have analytical solutions, it is possible to solve it numerically. This can be done straight-forwardly and requires fewer computational resources compared with modeling the full decohering state in a numerically truncated Hilbert space. Comparing with a numerically evolved state for small $|\alpha|$, we find that~\eqref{eq:homodyne:signal} corresponds exactly to the numerical solution. 

We plot the optical phase-space quadratures in Fig.~\ref{fig:homodyne} as a function of time for different values of the decay rate $\tilde{\kappa}_{\rm{c}}$, where we have assumed that the mechanical system is in the ground state with $|\beta| = 0$. We set the coupling to $\tilde{g}_0 = 1$, which means that the state should return to its starting point in phase-space at $\tau = 2\pi$ (after one mechanical oscillation).  We note that for unitary dynamics with $\tilde{\kappa}_{\rm{c}} = 0$ (the blue line), this is indeed what happens.  However, when $\tilde{\kappa}_{\rm{c}} \neq 0$, the trajectory slowly decays towards the vacuum state. We can prove this fact explicitly by examining the expression for $\braket{\hat a(\tau)}$  in~\eqref{eq:homodyne:signal}. The real value of the integral can be simplified and upper bounded, which allows us to prove that both of the real quantities $\hat X_{\rm{c}}(\tau)$ and $\hat P_{\rm{c}}(\tau)$ go to zero as $\tau \rightarrow \infty$, as expected. The proof can be found in Appendix~\ref{app:quadratures}. 

We also consider the intracavity optical quadratures when the mechanics is in the thermal state~\eqref{eq:thermal:state}. Since thermal states represent a weighted average over coherent states, we focus on the term in~\eqref{eq:homodyne:signal} that contains $\beta$. By integrating with the appropriate weighting  for the thermal state, we find 
\begin{equation}
\frac{1}{\bar{n}\pi} \int_{\mathbb{C}} \mathrm{d}^2 \beta \, e^{- |\beta|^2/\bar{n}} \, e^{G(\tau) \beta^* - G^*(\tau) \beta} = e^{- |G|^2 \, \bar{n}}.
\end{equation}
Inserting this into~\eqref{eq:homodyne:signal}, we find  the following expression for optical quadratures:
\begin{align} \label{eq:homodyne:signal:thermal}
&\braket{\hat a(\tau)}_{\mathrm{th}} =\alpha \, e^{|\alpha|^2 \left(e^{ - 2\, i \, A(\tau)} \, e^{- \tilde{\kappa}_{\rm{c}}  \tau}-1\right) } e^{-|G(\tau)|^2(1 + 2 \bar{n}) /2}\,   e^{- i\, A(\tau)}  \,  	\nonumber \\
&\quad\times e^{ -  \tilde{\kappa}_{\rm{c}} \tau/2} \, \mathrm{exp} \left[\tilde{\kappa}_{\rm{c}} \, |\alpha|^2 \int^\tau_0 \mathrm{d}\tau'  \, e^{- \tilde{\kappa}_{\rm{c}} \tau'} \,e^{-2\,i\,A(\tau')}  \, e^{i \, B(\tau', \tau) } \right]. 
\end{align}
We note that the quadratures tend to zero for large $\bar{n}$ and $|G|$. However, since $G$ is an oscillating function, the state still returns to its original value in phase space whenever $|G| = 0$, which occurs when there the optical and mechanical modes disentangle.

\subsection{Fidelity for generating optical cat-states}
For our final example we consider the generation of optical cat states of the intracavity field in the presence of optical loss. The cat states allow, among other things, for logical qubits to be encoded~\cite{cochrane1999macroscopically, leghtas2013hardware, mirrahimi2014dynamically} which makes them interesting for various information-processing schemes.

It has been shown that two initially coherent states [such as those shown in~\eqref{eq:initial:state}] evolve under the  Hamiltonian~\eqref{eq:Hamiltonian} as~\cite{bose1997preparation, mancini1997ponderomotive}
\begin{align}\label{non:linear:state:evolution}
\ket{\Psi(\tau)} =& \,    e^{-|\alpha|^2/2}\,\sum_{n = 0}^\infty  \frac{\alpha^n}{\sqrt{n!}} \,   e^{- i \,\left(F_{a}+  F_{+} \, F_{-}  \right)\, n^2}e^{ - i   \,  \mathrm{Im}\left(G^*\, \beta \right)\,  n} \, 
\nonumber \\
&\qquad\qquad\times \ket{n}_{\rm{c}} \otimes  \ket{e^{- i \tau} \, \beta + e^{- i \tau} \,G\, n}_{\rm{m}}  \, ,
\end{align}
where  $\ket{e^{- i \tau} \, \beta + e^{- i \tau} \,G \, n}_{\rm{m}}$ is a coherent state of the mechanics.~\footnote{Note that the notation used here is slightly different compared with that in~\cite{bose1997preparation}.}

When the optomechanical coupling is constant, with its rescaled form being $\tilde{g}(\tau) \equiv \tilde{g}_0 = g_0/\omega_{\mathrm{m}}$, we find that the optical and mechanical states evolve into  separable states at $\tau = 2\pi$.
We see this from the expressions for  $F_+$ and $F_-$ in~\eqref{eq:F:coefficients}, which become $F_+ = F_- = 0$ at $\tau = 2\pi$, which in turn implies that $G = 0$. The traced-out cavity state then becomes
\begin{align} \label{eq:pure:cat:state}
\ket{\Psi(2\pi) }_{\rm{c}}= e^{- |\alpha|^2/2} \sum_{n = 0}^\infty \frac{\alpha^n}{\sqrt{n!}} e^{2\pi \,i \,\tilde{g}_0^2 \, n^2} \ket{n}_{\rm{c}}. 
\end{align}
The value of $\tilde{g}_0$ determines the number of components of the cat state~\cite{bose1997preparation}. 
For example, $\tilde{g}_0 = \frac{1}{2}$ yields the  two-component cat state
\begin{equation}
\ket{\Psi( 2\pi )}_{\rm{c}}  =  \left( \frac{ 1 + i  }{2}\ket{+ \alpha} + \frac{1 - i}{2} \ket{- \alpha} \right), 
\end{equation}
where the distance between the two components in phase space is given by the coherent-state parameter $\alpha$. Three- and four-component cat states can be similarly generated with $\tilde{g}_0 = 1/\sqrt{6}$ and $\tilde{g}_0 = 1/2 \sqrt{2}$ ( see~\cite{bose1997preparation}).

To highlight the non-classical features of the state and how these are expected to decay with increasing $\tilde{\kappa}_{\rm{c}}$, we numerically evolve the state and compute its Wigner function $W(X,P)$ for various $\tilde{g}_0$ and $\tilde{\kappa}_{\rm{c}}$. The Wigner function for multicomponent cat states is shown in Figure~\ref{fig:Wigner} for $|\alpha| = \sqrt{3}$. Here the red areas denote the nonclassical features of the state, which can be seen to degrade as the state decoheres. However, since this computation relies on a truncated Hilbert space, we are not able to examine large $|\alpha|$. 

We proceed to derive an expression for the fidelity $\mathcal{F}$ of generating an optical cat state in the presence of optical decoherence. We do so by taking the overlap between the ideal cat state~\eqref{eq:pure:cat:state} and the noisy state evolving with $\hat{\mathcal{S}}(\tau)$ given  in~\eqref{eq:noisy:dynamics:general}. The vectorized final state is given by $\kket{\varrho(\tau)} = \hat{\mathcal{S}} (\tau)  \ket{\alpha}\ket{\beta} \otimes \ket{\alpha^*}\ket{\beta^*}$ and the overlap becomes  
\begin{equation}
\mathcal{F}= \bra{\Psi(2\pi)} \hat \varrho(2\pi) \ket{\Psi(2\pi)} = \bbraket{\Psi^\dag(2\pi) | \varrho (2\pi)},
\end{equation}
where $\bbra{\Psi^\dag(2\pi)}$ is the vectorized ideal cat state~\eqref{eq:pure:cat:state}. We find the following expression for the fidelity (see Appendix~\ref{app:fidelity} for the full calculation):
\begin{align} \label{eq:fidelity}
\mathcal{F} &=  e^{- 2|\alpha|^2} \sum_{n= 0}^\infty \sum_{n'= 0}^\infty \frac{|\alpha|^{2(n+n')} }{n!n'!}  e^{- \tilde{\kappa}_{\rm{c}} \pi (n+n')}  \\
&\times \mathrm{exp} \left[\tilde{\kappa}_{\rm{c}}\, |\alpha|^2 \int^{2\pi}_0 \mathrm{d}\tau' e^{- \tilde{\kappa}_{\rm{c}} \tau'} \, e^{- 2 \, i \, A(\tau')(n-n')}\right]. \nonumber
\end{align}
Setting $\tilde{\kappa}_{\mathrm{c}}=  0$, we recover $\mathcal{F} = 1$, as expected. For nonzero $\tilde{\kappa}_{\mathrm{c}}$, we find that~\eqref{eq:fidelity} corresponds exactly to numerical results. 

The expression~\eqref{eq:fidelity} can be simplified further. In Appendix~\ref{app:fidelity}, we show how~\eqref{eq:fidelity} can be expanded in  increasing orders of  $A(\tau)$ and $\tilde{\kappa}_{\rm{c}} |\alpha|^2$. The somewhat lengthy result is shown in~\eqref{app:eq:simplified:fidelity}. While formally infinite, the expression indicates a complicated relationship between $\tilde{\kappa}_{\rm{c}}$, $\alpha$ and $\tilde{g}_0 $ [note that $A(\tau)\propto \tilde{g}_0^2$]. The advantage of the expression~\eqref{app:eq:simplified:fidelity} is that when either $\tilde{g}_0 \ll1$ or $\tilde{\kappa}_{\rm{c}} |\alpha|^2 \ll 1$, the fidelity can be straight-forwardly expanded and evaluated to the desired order.

To obtain a more intuitive limit of the fidelity, we proceed to bound $\mathcal{F}$ from above and below. We find (see Appendix~\ref{app:bounding:fidelity} for the full calculation):
\begin{align} \label{eq:fidelity:bounds}
2 \, e^{- 2 \, |\alpha|^2} \mathrm{sh} + e^{- |\alpha|^2 ( 1 + e^{- \pi \tilde{\kappa}_{\rm{c}}})^2 } \leq \mathcal{F} \leq e^{- |\alpha|^2 ( 1 - e^{- \pi \tilde{\kappa}_{\rm{c}}}) ^2 }, 
\end{align}
where $\mathrm{sh}:= \sinh (2 \,|\alpha|^2 e^{- \pi \tilde{\kappa}_{\rm{c}}}) $.  

We plot $\mathcal{F}$ and its upper and lower bounds~\eqref{eq:fidelity:bounds} as a function of $\tilde{\kappa}_{\rm{c}}$ for different values of $\alpha$ in Fig.~\ref{fig:fidelity}. The shaded areas indicate the upper and lower bounds in~\eqref{eq:fidelity:bounds}. We note that $\mathcal{F}$ rapidly decreases with $\tilde{\kappa}_{\rm{c}}$ for higher values of $|\alpha|$. We also note that a coherent state with $|\alpha| = 1$ retains a fairly high fidelity, which is due to the large non-zero overlap between $\ket{\alpha = 1}$ and the vacuum  $\ket{0}$. 

 The upper bound~\eqref{eq:fidelity:bounds} allows us to bound the decay rate $\tilde{\kappa}_{\rm{c}}$ given a desired fidelity. As an example, let us consider the case where we wish to prepare a two-component optical cat state with $\tilde{g}_0 = 0.5$ using a coherent state with $|\alpha|^2 = 10$. If we wish to generate the cat state with a fidelity of $\mathcal{F} = 0.99$, we find that we require roughly $\tilde{\kappa}_{\rm{c}} \sim 0.01$. The linewidth of a cavity  is given by the angular frequency $\kappa_{\rm{c}} = \pi c/2 L F$~\cite{hunger2010fiber}, where  $c$ is the speed of light, $L$ is the cavity length, and $F$ is the cavity finesse. Given a cavity of length $L = 10$\,mm and a finesse of $F = $ 500\,000, we find $\kappa_{\rm{c}}/(2\pi) =  15$\,kHz. We thus require a mechanical frequency of $\omega_{\rm{m}} /2 \pi=1.5$\,MHz, such that $\tilde{\kappa}_{\rm{c}} = \kappa_{\rm{c}}/\omega_{\rm{m}} =  0.01$, and a coupling strength of $g_0/2\pi  =0.75$\,MHz, such that $\tilde{g}_0 = g_0/\omega_{\rm{m}} =  0.5$. While a finesse, linewidth, and mechanical frequency of similar magnitude have been demonstrated experimentally~\cite{de2020strong,pontin2020ultranarrow},  a single-photon coupling of this strength has not yet been achieved. To access the intracavity cat state, we envision the utilization of a scheme that coherently opens the cavity, such as that proposed by~\citet{tufarelli2014coherently}.

\section{Summary and outlook} \label{sec:conclusions}
In this work we solved the Lindblad master equation for optical decoherence in a nonlinear optomechanical system. The solution involved vectorizing the Lindblad equation as well as partitioning the nonunitary time evolution into treatable contributions. Our main result, shown in~\eqref{eq:noisy:dynamics:general}, is a compact expression that encodes the full nonunitary evolution of the optical and mechanical states. 
To demonstrate the applicability of our method, we derived the fidelity for preparing optical cat states with a leaking cavity. The resulting expressions allowed us to bound the optical decay rate required to produce cat states at a desired fidelity.  

Our method opens up the possibility for considering optical decoherence in a variety of contexts, such as proposals for generating macroscopic superpositions~\cite{bose1999scheme, marshall2003towards, kleckner2008creating}, and sensing schemes~\cite{schneiter2020optimal, qvarfort2021optimal}. Potentially, the method could be used to provide a theoretical description of the regime of large thermal motion and weak single-photon coupling~\cite{brawley2016nonlinear,leijssen2017nonlinear}; however, we note that it does not yet include a drive of the cavity field or an input-output formalism, both of which are fundamental to many experimental setups. We also note that while some of the results presented here were given in closed-form expressions, such as the optical quadratures~\eqref{eq:homodyne:signal}, other properties of the system, such as the number of phonons of the mechanical state, must be studied perturbatively by expanding the expressions for small $\tilde{\kappa}_{\mathrm{c}}$. We also note that once the mechanical coupling is of  strength comparable to the mechanical frequency, one can no longer consider optical and mechanical decoherence separately~\cite{hu2015quantum}. We leave these considerations to future work. Finally, we also note that our method applies to any system that exhibits dynamics captured by the nonlinear Hamiltonian~\eqref{eq:Hamiltonian}, such as electro-optical systems~\cite{tsang2010cavity}.

\section*{Acknowledgments}
We  thank Lindsay Orr, Suocheng Zhao, Jack Clarke, Daniel Goldwater, Ying Lia Li, Dennis R\"{a}tzel, Marko Toro\v{s}, Doug Plato,  Daniel Braun, Igor Pikovski, Myungshik Kim, Ivette Fuentes, Alessio Serafini,   Tania Monteiro, Andr\'{e} Xuereb, Anja Metelmann  and Sougato Bose for helpful discussions. We also thank the referees for their careful reading of the manuscript, which helped us improve it. S.Q.~was supported by an Engineering and Physical Sciences Research Council Doctoral Prize Fellowship.

\section*{Data availability statement}
The code used to generate the Wigner function plot (Fig.~\ref{fig:Wigner}) can be found in the following GitHub repository: \href{https://github.com/sqvarfort/noisy-optical-cat-states}{https://github.com/sqvarfort/noisy-optical-cat-states}. 

\bibliographystyle{apsrev4-2}
\bibliography{cat_states}

\onecolumngrid

\newpage

\appendix

\widetext

\section{Derivation of the time-evolution partition}  \label{app:interaction:picture}
In this appendix, we provide a derivation of the time partitioning of the evolution operators, which we make frequent use of in the main text. We start by considering unitary dynamics, but then we generalize the results to nonunitary dynamics in the vectorized language. 

\subsection{Unitary dynamics}  \label{app:unitary:interaction:picture}
This proof follows the usual derivation of the interaction picture, however we show that any partition of the Hamiltonian will do, even one where the two parts are time-dependent. 

We start by writing down the Hamiltonian in the Schr\"{o}dinger picture. Henceforth, all  $S$ indices denote a quantity in the Schr\"{o}dinger picture, while $I$ indices denote the interaction picture. We assume that the Hamiltonian is given by 
\begin{equation}
\hat H_{\mathrm{S}}(t) = \hat H_{0, \mathrm{S}}(t) + \hat H_{1, \mathrm{S}}(t). 
\end{equation}
In the standard derivation, $\hat H_{0, \mathrm{S}}(t)$ is often taken to the free (time-independent) evolution of the system. While this is often a convenient choice, it is not always necessary. 

We proceed to write down the Schr\"{o}dinger equation in the Schr\"{o}dinger picture:
\begin{equation} \label{app:eq:SE}
i\hbar \frac{\mathrm{d}}{\mathrm{d}t} \ket{\psi(t)}_{\mathrm{S}}= \left[ \hat H_{0, \mathrm{S}}(t) +\hat H_{1, \mathrm{S}}(t) \right] \ket{\psi(t)}_{\mathrm{S}}.
\end{equation}
Next, we define the state in the frame that we wish to study:
\begin{equation} \label{app:eq:unitary:interaction:picture:state}
\ket{\psi(t)}_{\mathrm{I}} = \hat U_0^\dag(t) \ket{\psi(t)}_{\mathrm{S}}, 
\end{equation}
where  $\hat U_0(t)$ is given by the time-ordered integral
\begin{equation}
\hat U_0(t) = \overleftarrow{\mathcal{T}} \mathrm{exp}\left[ - \frac{i}{\hbar} \int^t_0 \mathrm{d}t' \, \hat H_{0, \mathrm{S}}(t') \right].
\end{equation}
Now, we examine the Schrödinger equation in the interaction picture. We find 
\begin{equation}
i \hbar \frac{\mathrm{d}}{\mathrm{d}t} \ket{\psi(t)}_{\mathrm{I}} = \frac{\mathrm{d}}{\mathrm{d}t} \hat U_0^\dag(t) \ket{\psi(t)}_{\mathrm{S}} + \hat U_0^\dag(t) \frac{\mathrm{d}}{\mathrm{d}t} \ket{\psi(t)} _{\mathrm{S}}.
\end{equation}
However, the derivative of the state appears in the Schr\"{o}dinger equation. Inserting the right-hand side of~\eqref{app:eq:SE}, we find 
\begin{equation}
i \hbar \frac{\mathrm{d}}{\mathrm{d}t} \ket{\psi(t)}_{\mathrm{I}} = \frac{\mathrm{d}}{\mathrm{d}t} \hat U_0^\dag(t) \ket{\psi(t)}_{\mathrm{S}} -  \frac{i}{\hbar} \hat U_0^\dag (t) \left[ \hat H_{0, \mathrm{S}}(t) +\hat H_{1, \mathrm{S}}(t) \right] \ket{\psi(t)}_{\mathrm{S}}. 
\end{equation}
The derivative of  $\hat U^\dag_0(t)$ is given by
\begin{equation}
\frac{\mathrm{d}}{\mathrm{d}t} \hat U_0^\dag (t) = \left( \frac{\mathrm{d}}{\mathrm{d}t} \hat U_0(t) \right)^\dag = \left( - \frac{i}{\hbar} \hat H_{0, \mathrm{S}}(t) \, \hat U_0(t) \right)^\dag = \frac{i}{\hbar} \hat U_0^\dag (t) \, \hat H_{0, \mathrm{S}}(t),
\end{equation}
which follows from Leibniz's integral rule and the fact that the adjoint operation commutes with the derivative. Inserting this into the expression, we find that 
\begin{equation} \label{app:eq:interaction:picture:SE}
i \hbar \frac{\mathrm{d}}{\mathrm{d}t} \ket{\psi(t)}_{\mathrm{I}} = \frac{i}{\hbar}  \hat U_0^\dag (t) \, \hat H_{0, \mathrm{S}}(t)\ket{\psi(t)}_{\mathrm{S}} -  \frac{i}{\hbar} \hat U_0^\dag (t) \left[ \hat H_{0, \mathrm{S}}(t) +\hat H_{1, \mathrm{S}}(t) \right] \ket{\psi}_{\mathrm{S}} = \hat H_{1, \mathrm{I}}(t) \,  \ket{\psi(t)}_{\mathrm{I}}, 
\end{equation}
where we have defined the second Hamiltonian term in the interaction picture as 
\begin{equation}
\hat H_{1, \mathrm{I}}(t) = \hat U_0^\dag (t) \, \hat H_{1,\mathrm{S}}(t) \, \hat U_0(t) . 
\end{equation}
The formal solution to~\eqref{app:eq:interaction:picture:SE} is 
\begin{equation}
\ket{\psi(t)}_{\mathrm{I}} = \hat U_1(t) \ket{\psi(0)}_{\mathrm{I}}, 
\end{equation}
where $\hat U_1(t)$ is defined as
\begin{equation}
\hat U_1(t) = \overleftarrow{\mathcal{T}} \mathrm{exp}\left[ - \frac{i}{\hbar} \int^t_0 \mathrm{d}t' \, \hat U_0^\dag (t) \, \hat H_{1,\mathrm{S}}(t) \, \hat U_0(t) \right].
\end{equation}
Then, we revert to the state in the Schr\"{o}dinger picture. Multiplying~\eqref{app:eq:unitary:interaction:picture:state} by $\hat U_0(t)$ on both sides, we find
\begin{equation}
\ket{\psi(t)}_{\mathrm{S}} = \hat U_0(t) \ket{\psi(t)}_{\mathrm{I}} = \hat U_0(t) \, \hat U_1(t) \ket{\psi(0)}.
\end{equation}
Thus, we have shown that any arbitrary partition of the Hamiltonian, and thereby the time-evolution, allows us to pairwise evaluate the time-evolution operators. 

\subsection{Nonunitary dynamics}  \label{app:non:unitary:interaction:picture}

This proof follows the derivation of the evolution in the interaction picture used in the preceding section.  Let the Lindbladian be given by
\begin{equation}
\hat{\mathcal{L}}(t) = \hat{\mathcal{L}}_0(t) + \hat{\mathcal{L}}_1(t).
\end{equation}
Then, the vectorized Lindblad equation in the Schr\"{o}dinger picture, which we denote by the subscript ${\mathrm{S}}$, is (see the main text for the derivation)
\begin{equation} \label{eq:lindblad}
\frac{\mathrm{d}}{\mathrm{d}t} \kket{\varrho(t)}_{\mathrm{S}}= \left( \hat{\mathcal{L}}_{0,\mathrm{S}}(t) + \hat{\mathcal{L}}_{1,\mathrm{S}}(t) \right) \kket{\varrho(t)}_{\mathrm{S}}. 
\end{equation}
We define the new state in the interaction picture (which we similarly denote with $\mathrm{I}$)
\begin{equation} \label{app:eq:vectorized:interaction:picture}
\kket{\varrho(t)}_{\mathrm{I}} = \hat{\mathcal{S}}_0(t)^{-1} \kket{\varrho(t)}_{\mathrm{S}}, 
\end{equation}
where $\hat{\mathcal{S}}_0(t)$ is defined as
\begin{equation}
\hat{\mathcal{S}}_0(t) = \overleftarrow{\mathcal{T}} \mathrm{exp}\left[ \int^t_0 \mathrm{d}t' \, \hat{\mathcal{L}}_0(t') \right].
\end{equation}
Then we examine the Lindblad equation for the vectorized state, shown in~\eqref{eq:vectorized:Lindblad}. Taking the derivative, we find 
\begin{align}
\frac{\mathrm{d}}{\mathrm{d}t} \kket{\varrho(t)}_{\mathrm{I}} &= \frac{\mathrm{d}}{\mathrm{d}t} \left[ \hat{\mathcal{S}}_0^{-1}(t) \kket{\varrho(t)}_{\mathrm{S}} \right] = \frac{\mathrm{d}}{\mathrm{d}t} \hat{\mathcal{S}}_0^{-1}(t) \kket{\varrho(t)}_{\mathrm{S}}+ \hat{\mathcal{S}}_0^{-1}(t) \frac{\mathrm{d}}{\mathrm{d}t} \kket{\varrho(t)}_{\mathrm{S}}. 
\end{align}
We now have to consider the derivative of an inverted matrix. Using the identity $\mathds{1} = \hat{\mathcal{S}}_0^{-1}(t) \, \hat{\mathcal{S}}_0(t)$, which is well defined for the exponential map, we find
\begin{equation}
\frac{\mathrm{d}}{\mathrm{d}t}\mathds{1} = 0 = \frac{\mathrm{d}}{\mathrm{d}t} \hat{\mathcal{S}}_0^{-1}(t) \, \hat{\mathcal{S}}_0(t) + \hat{\mathcal{S}}_0^{-1}(t) \, \frac{\mathrm{d}}{\mathrm{d}t}\hat{\mathcal{S}}_0(t). 
\end{equation}
Rearranging, we have that 
\begin{equation}
\frac{\mathrm{d}}{\mathrm{d}t} \hat{\mathcal{S}}_0^{-1}(t) = - \hat{\mathcal{S}}_0^{-1}(t) \, \frac{\mathrm{d}}{\mathrm{d}t}\hat{\mathcal{S}}_0(t) \,  \hat{\mathcal{S}}_0^{-1}(t). 
\end{equation}
The derivative of $ \hat{\mathcal{S}}_0(t) $ again follows from Leibniz's integral rule:
\begin{align}
\frac{\mathrm{d}}{\mathrm{d}t}  \hat{\mathcal{S}}_0(t)  = \hat{\mathcal{L}}_{0, \mathrm{S}}(t) \, \hat{\mathcal{S}}_0(t) .
\end{align}
Thus we find 
\begin{equation}
\frac{\mathrm{d}}{\mathrm{d}t} \hat{\mathcal{S}}_0^{-1}(t) = - \hat{\mathcal{S}}_0^{-1}(t) \, \hat{\mathcal{L}}_{0, \mathrm{S}}(t). 
\end{equation}
We then note that the derivative of the vectorized state is given by the Lindblad equation~\eqref{eq:lindblad}. 
Inserting this, we find 
\begin{align} \label{eq:manipulated}
\frac{\mathrm{d}}{\mathrm{d}t} \kket{\varrho(t)}_{\mathrm{I}} &=  -\hat{\mathcal{S}}_0^{-1}(t)  \, \hat{\mathcal{L}}_{0,{\mathrm{S}}} \,  \kket{\varrho(t)}_{\mathrm{S}} + \hat{\mathcal{S}}_0^{-1}(t) \,  \left( \hat{\mathcal{L}}_{0,{\mathrm{S}}} + \hat{\mathcal{L}}_{1,{\mathrm{S}}} \right) \kket{\varrho(t)}_{\mathrm{S}}. 
\end{align}
Canceling the terms and defining 
\begin{equation}
\hat{\mathcal{L}}_{1, {\mathrm{I}}} (t)= e^{- \hat{\mathcal{L}}_{0,{\mathrm{S}}}\, t} \,  \hat{\mathcal{L}}_{1,{\mathrm{S}}}  \,  e^{ \hat{\mathcal{L}}_{0,{\mathrm{S}}} \, t}, 
\end{equation}
we have 
\begin{align}
\frac{\mathrm{d}}{\mathrm{d}t} \kket{\varrho(t)}_{\mathrm{I}} &=   \hat{\mathcal{L}}_{1, I}(t) \kket{\varrho(t)}_{{\mathrm{I}}}. 
\end{align}
The formal solution to the equation is 
\begin{equation}
\kket{\varrho(t)}_{\mathrm{I}} = \hat{\mathcal{S}}_1 (t) \, \kket{\varrho(0)}_{\mathrm{I}}, 
\end{equation}
where $\hat{\mathcal{S}}_1$ is defined as
\begin{equation}
\hat{\mathcal{S}}_1(t) =   \overleftarrow{\mathcal{T}}\mathrm{exp} \left[ \int^t_0 \mathrm{d}t' \,e^{- \hat{\mathcal{L}}_{0,{\mathrm{S}}} \, t'} \, \hat{\mathcal{L}}_{1,{\mathrm{S}}}(t')  \,  e^{ \hat{\mathcal{L}}_{0,{\mathrm{S}}} \, t'} \right].
\end{equation}
Finally, we wish to return to the Schr\"{o}dinger picture. Multiplying~\eqref{app:eq:vectorized:interaction:picture} on both sides by $\hat{\mathcal{S}}_0(t)$, we find 
\begin{equation}
\kket{\varrho(t)}_{\mathrm{S}} = \hat{\mathcal{S}}_0(t)   \kket{\varrho(t)}_{\mathrm{I}} = \hat{\mathcal{S}}_0(t) \, \hat{\mathcal{S}}_1(t)  \kket{\varrho(0)}.
\end{equation}
This proves that the nonunitary time evolution can  be partitioned into two arbitrary terms that evolve with different parts of $\hat{\mathcal{S}}(t)$, just as is the case for unitary dynamics. Note that we have not made any assumptions about the Hermiticity of either $\hat{\mathcal{S}}_0(t)$ or $\hat{\mathcal{S}}_1(t)$, but that everything follows from the fact that the inverse is well defined for exponential maps.

\section{Expanding the exponential} \label{app:expanding:the:exponential}
In this appendix, we derive a simpler form of the integral  in~\eqref{eq:explicit:main:result}. This will greatly help us in subsequent calculations. The integral is given by 
\begin{align} \label{app:eq:integral}
\overleftarrow{\mathcal{T}} \mathrm{exp} \left[\tilde{\kappa}_{\rm{c}} \int^\tau_0 \mathrm{d}\tau'  \, e^{- \tilde{\kappa}_{\rm{c}} \tau'} \,e^{-2\,i\,(F_a+F_+ F_- )\,\hat{N}_a}\,e^{-i\,F_+\,\hat{B}_+}\,e^{-i\,F_-\,\hat{B}_-}  \,  \hat a \otimes e^{2\,i\,(F_a+F_+ F_- )\,\hat{N}_a}\,e^{i\,F_+\,\hat{B}_+}\,e^{i\,F_-\,\hat{B}_-}  \, \hat a \,  \right].
\end{align}
We start by writing~\eqref{app:eq:integral} as a Neumann series:
\begin{align} \label{app:eq:expanded:integral}
&\overleftarrow{\mathcal{T}} \mathrm{exp}\left[ \tilde{\kappa}_{\rm{c}} \int^\tau_0 \mathrm{d}\tau'  \, e^{- \tilde{\kappa}_{\rm{c}} \tau'} \, e^{-2\,i\, A (\tau')\,\hat{N}_a}\,  \hat a\,\hat D(G(\tau'))    \otimes  e^{2 \,i \,A(\tau')\hat N_a}  \, \hat a\,   \hat D(G^*(\tau')) \, \right] \nonumber \\
&= 1  + \tilde{\kappa}_{\rm{c}} \int^\tau_0 \mathrm{d}\tau'  \, e^{- \tilde{\kappa}_{\rm{c}} \tau'} \, e^{-2\,i \,A (\tau')\,\hat{N}_a}\,  \hat a\,\hat D(G(\tau'))    \otimes  e^{2\, i \,A(\tau')\hat N_a} \, \hat a \,   \hat D(G^*(\tau')) \, \nonumber \\
&\qquad+ \tilde{\kappa}_{\rm{c}}^2 \int^\tau_0 \mathrm{d}\tau' \, \int^{\tau'}_0 \mathrm{d}\tau'' \, e^{- \tilde{\kappa}_{\rm{c}}(\tau' + \tau'') }\, e^{-2\,i \,A (\tau')\,\hat{N}_a}\,  \hat a \, e^{-2\,i \,A (t'')\,\hat{N}_a}\,  \hat a \,  \hat D(G(\tau')) \hat D(G(\tau'')) \nonumber \\
&\qquad\qquad\qquad\qquad \otimes  e^{ 2\,i\, A(\tau')\hat N_a} \,  \hat a \, e^{2\, i \,A(\tau'')\hat N_a} \,  \hat a \, \hat D(G^*(\tau''))\,\hat D(G^*(\tau''))  + \cdots,
\end{align}
where higher order terms show a significant increase in complexity.
 
We wish to simplify the expression in~\eqref{app:eq:expanded:integral} by collecting the $\hat a$ operators. We use the fact that $e^{ i X \hat N_a }\,  \hat a \,e^{-i X \hat N_a} = e^{- i X }\,\hat a$, which allows us to conclude that swapping an $\hat a$ operator with an exponential of $\hat N_a$ generates an extra phase:
\begin{equation}
 \hat a \,e^{-i X \hat N_a} =e^{- i X }\,e^{- i X \hat N_a } \, \hat a. 
\end{equation}
For example, consider the second-order term in the expression~\eqref{app:eq:expanded:integral}. Starting with the left-hand mode that arose from the vectorization, we find
\begin{align}
e^{-iX_1 \hat N_a } \,\hat a \, e^{-iX_2 \hat N_a} \hat a &= e^{- i X_1 \hat N_a} e^{ - i X_2 \hat N_a} \, e^{i X_2 \hat N_a } \, \hat a \, e^{- i X_2 \hat N_a} \hat a \nonumber \\
&= e^{- i X_1 \hat N_a} e^{ - i X_2 \hat N_a} \,  e^{- i X_2 } \, \hat a^2. 
\end{align}
Similarly, for the third order expression, we find
\begin{align}
e^{-iX_1 \hat N_a } \hat a \, e^{-iX_2 \hat N_a} \hat a \, e^{- i X_3 \hat N_a} \hat a & = e^{-iX_1 \hat N_a } \hat a \, e^{-iX_2 \hat N_a}  \, e^{- i X_3 \hat N_a} e^{- i X_3} \hat a^2 \nonumber \\
&= e^{-iX_1 \hat N_a }  \, e^{-iX_2 \hat N_a}  \, e^{- i X_2}  \hat a\, e^{- i X_3 \hat N_a} e^{- i X_3} \hat a^2 \nonumber \\
&=  e^{-iX_1 \hat N_a }  \, e^{-iX_2 \hat N_a} \, e^{- i X_3 \hat N_a}  \, e^{- i X_2}  e^{- 2i X_3}  \hat a^3. 
\end{align}
The general formula for order $n$ reads:
\begin{align}
\prod_{j= 1}^n \left( e^{- i X_j \hat N_a} \,  \hat a \right) &= e^{- i X_1 \hat N_a} \, \hat a \, e^{- i X_2 \hat N_a} \, \hat a \, e^{- i X_3 \hat N_a } \, \hat a \cdots e^{- i X_n \hat N_a } \, \hat a \nonumber \\
&= e^{- i X_1 } \, e^{- i X_2 } \, e^{- 2 i X_3 } \cdots e^{- (n-1) i X_n } \, e^{- i X_1 \hat N_a} \, e^{- i X_2 \hat N_a} \, e^{- i X_3 \hat N_a} \cdots e^{- i X_n \hat N_a} \, \hat a^n . 
\end{align}
The same can be done for the right-hand-mode terms. 

Inserting the result into~\eqref{app:eq:expanded:integral}, we find that the phases $e^{- iX_j }$ from the left-hand mode cancel with those from the right-hand side, which  look like $e^{i X_j}$. We are left with 
\begin{align} \label{app:eq:simplified:exponential}
\overleftarrow{\mathcal{T}} \mathrm{exp}&\left[ \tilde{\kappa}_{\rm{c}} \int^\tau_0 \mathrm{d}\tau'  \, e^{- \tilde{\kappa}_{\rm{c}} \tau'} \, e^{-2\,i \,A (\tau')\,\hat{N}_a} \,  \hat a\,\hat D(G(\tau))   \otimes  e^{2\, i \,A(\tau')\hat N_a} \, \hat a\,   \hat D(G^*(\tau'))  \, \right] \nonumber \\
&= 1  + \tilde{\kappa}_{\rm{c}} \int^\tau_0 \mathrm{d}\tau'  \, e^{- \tilde{\kappa}_{\rm{c}} \tau'} \, e^{-2\,i \,A (\tau')\,\hat{N}_a} \,  \hat a \,\hat D(G(\tau'))  \otimes  e^{2\, i \,A(\tau')\hat N_a} \, \hat a\,   \hat D(G^*(\tau'))  \, \nonumber \\
&\quad+ \tilde{\kappa}_{\rm{c}}^2 \int^\tau_0 \mathrm{d}\tau' \, \int^{\tau'}_0 \mathrm{d}\tau'' \, e^{- \tilde{\kappa}_{\rm{c}}(\tau' + \tau'')} \, e^{-2\,i \,A (\tau')\,\hat{N}_a}\,  e^{-2\,i \,A (\tau'')\,\hat{N}_a}\, \hat a ^2 \,  \hat D(G(\tau')) \, \hat D(G(\tau''))  \nonumber \\
&\qquad \qquad \qquad\qquad  \otimes  e^{2\, i\, A(\tau')\hat N_a} \,  e^{ 2\,i\, A(\tau'')\hat N_a} \,  \hat a^2 \, \hat D(G^*(\tau'))\,\hat D(G^*(\tau''))  + \cdots . 
\end{align}
This expression will be frequently used in the following appendixes. 

\section{Derivation of the photon number expectation value} \label{app:photon:number}
We proceed by computing the expectation value of $\hat N_a = \hat a^\dag \hat a$ as a function of time $\tau$. In the vectorized language, it is given by $\braket{\hat N_a(\tau)} := \mathrm{Tr} \bigl[ \hat N_a \, \hat \varrho(\tau) \bigr] = \bbraket{\hat N_a | \varrho(\tau)}$, where $\kket{\varrho(\tau)} = \hat{\mathcal{S}}(\tau) \kket{\varrho_0}$, with $\kket{\varrho_0}$ being the initial vectorized state. 
The operator $\hat N_a$ when vectorized is given by
\begin{align}
\kket{\hat N_a } = \sum_{n = 0}^\infty \, \sum_{m = 0}^\infty \, n \, \ket{n}\ket{m}\otimes \ket{n}  \ket{m}. 
\end{align}
Here, we have included an identity operator in the vectorization that acts on the mechanical subsystem. The tensor product will consistently refer to the separation of the left-hand and right-hand modes of the vectorization. 

To derive $\braket{\hat N_a (\tau)}$, we start from the full expression for $\hat{\mathcal{S}}(\tau)$, which reads
\begin{align}
\hat{\mathcal{S}}(\tau) &= \left( e^{- i \,\hat N_b\, \tau} \, e^{- i \,F_a \, \hat N_a^2} \, e^{- i \,F_+ \hat N_a \, \hat B_+} \, e^{- i \,F_- \hat N_a \, \hat B_-}e^{- \tilde{\kappa}_{\rm{c}} \tau \hat N_a/2 }\right)   \otimes \left( e^{i \,\hat N_b\, \tau} \, e^{ i \,F_a \, \hat N_a^2} \, e^{ i \,F_+ \hat N_a \, \hat B_+} \, e^{ i \,F_- \hat N_a \, \hat B_-}e^{-  \tilde{\kappa}_{\rm{c}} \tau \hat N_a/2 }\right)   \nonumber \\
&\quad \times \overleftarrow{\mathcal{T}} \mathrm{exp} \left[\tilde{\kappa}_{\rm{c}}  \int^\tau_0 \mathrm{d}\tau'  \, e^{- \tilde{\kappa}_{\rm{c}} \tau'} \,e^{-2\,i\,A(\tau')\,\hat{N}_a}\,  \hat a\,\hat D(G(\tau))  \otimes  e^{2\, i\, A(\tau') \hat N_a}\, \hat a \,   \hat D(G^*(\tau'))  \,  \right], 
\end{align}
together with the expressions of the initial coherent states of the optical field and mechanical element
\begin{align}
\kket{\varrho_0} = \ket{\alpha} \ket{\beta}\otimes \ket{\alpha^*} \ket{\beta^*}.
\end{align}
Then, we can compute the photon-number expectation value through the expression
\begin{align}
\braket{\hat N_a(\tau)} &= \sum_{n = 0}^\infty \sum_{m = 0}^\infty  n \, \bra{n}\bra{m} \otimes \bra{n}\bra{m}\left( e^{- i \, \hat N_b\, \tau} \, e^{- i  \, F_a \, \hat N_a^2} \, e^{- i \, F_+ \hat N_a \, \hat B_+} \, e^{- i \, F_- \hat N_a \, \hat B_-}e^{- \tilde{\kappa}_{\rm{c}} \tau \hat N_a/2 }\right)  \nonumber \\
&\qquad \qquad \qquad \qquad\qquad \qquad  \otimes \left( e^{ i \, \hat N_b\, \tau} \, e^{ i \, F_a \, \hat N_a^2} \, e^{ i \, F_+ \hat N_a \, \hat B_+} \, e^{ i \, F_- \hat N_a \, \hat B_-}e^{- \tilde{\kappa}_{\rm{c}} \tau \hat N_a/2 }\right)   \nonumber \\
&\quad \times \overleftarrow{\mathcal{T}} \mathrm{exp} \left[\tilde{\kappa}_{\rm{c}}  \int^\tau_0 \mathrm{d}\tau'  \, e^{- \tilde{\kappa}_{\rm{c}} \tau'} \,e^{-2\,i\,A(\tau')\,\hat{N}_a} \,  \hat a\,\hat D(G(\tau'))   \otimes  e^{2\, i \,A(\tau') \hat N_a}\, \hat a  \,   \hat D(G^*(\tau')) \,  \right]   \ket{\alpha} \ket{\beta}\otimes \ket{\alpha^*} \ket{\beta^*}. 
\end{align}
Applying the mechanical Fock states on the number operators $\hat N_b$ from the left, we note that the first phases with $e^{- i \, \hat N_b \tau}$ and $e^{i \, \hat N_b \tau}$ cancel. The same happens for the phases $e^{- i \, F_a \, \hat N_a^2}$ and $e^{i \, F_a \, \hat N_a^2}$. We are left with 
\begin{align}
\braket{\hat N_a(\tau)} &= \sum_{n = 0}^\infty \sum_{m = 0}^\infty n \, e^{-  \tilde{\kappa}_{\rm{c}} \tau \, n}\,  \bra{n}\bra{m} \otimes \bra{n}\bra{m} \left( e^{- i\, F_+n \, \hat B_+} \, e^{- i \, F_- n \, \hat B_-} \right)   \otimes \left( e^{i \, F_+ n \, \hat B_+}  \, e^{i \, F_-n \, \hat B_-}  \,  \right) \nonumber \\
&\quad \times \overleftarrow{\mathcal{T}} \mathrm{exp} \left[\tilde{\kappa}_{\rm{c}}  \int^\tau_0 \mathrm{d}\tau'  \, e^{- \tilde{\kappa}_{\rm{c}} \tau'} \,e^{-2\,i\,A(\tau')\,\hat{N}_a} \,  \hat a\,\hat D(G(\tau'))   \otimes  e^{2 \, i  \, A(\tau') \hat N_a}\, \hat a \,   \hat D(G^*(\tau'))  \,  \right]  \ket{\alpha} \ket{\beta}\otimes \ket{\alpha^*} \ket{\beta^*}. 
\end{align}
We then rewrite the exponential in terms of the expanded but simplified expression in~\eqref{app:eq:simplified:exponential}. Showing terms to second order, we find 
\begin{align}
\braket{\hat N_a(\tau)} &= \sum_{n= 0}^\infty \sum_{m = 0 }^\infty n \, e^{-  \tilde{\kappa}_{\rm{c}} \tau \,  n}\,  \bra{n}\bra{m} \otimes \bra{n}\bra{m} \left( e^{- i \, F_+n \, \hat B_+} \, e^{- i \, F_- n \, \hat B_-} \right)   \otimes \left( e^{i \,  F_+ n \, \hat B_+}  \, e^{i \, F_-n \, \hat B_-} \,    \right) \nonumber \\
&\quad\times  \biggl[ 1  + \tilde{\kappa}_{\rm{c}} \int^\tau_0 \mathrm{d}\tau'  \, e^{- \tilde{\kappa}_{\rm{c}} \tau'} \, e^{-2 \, i \, A (\tau')\,\hat N_a}\,  \hat a\,\hat D(G(\tau'))    \otimes  e^{ 2 \, i\, A(\tau')\, \hat N_a} \, \hat a \,  \hat D(G^*(\tau')) \,  \nonumber \\
&\quad\quad + \tilde{\kappa}_{\rm{c}}^2 \int^\tau_0 \mathrm{d}\tau' \, \int^{\tau'}_0 \mathrm{d}\tau'' \, e^{- \tilde{\kappa}_{\rm{c}}(\tau' + \tau'')} \, e^{-2 \, i \, [A (\tau') + A(\tau'')]\,\hat{N}_a}\, \hat a ^2 \,  \hat D(G(\tau') + G(\tau''))  \nonumber \\
&\qquad \qquad \qquad \qquad \otimes  e^{ 2 \,i \, [A(\tau') + A(\tau'')]\hat N_a} \,  \hat a^2 \, \hat D(G^*(\tau') + G^*(\tau''))  + \cdots   \biggr]  \ket{\alpha} \ket{\beta}\otimes \ket{\alpha^*} \ket{\beta^*}+\cdots, 
\end{align}
where we have used the fact that $\hat D(\gamma) \, \hat D(\xi) = e^{( \gamma \xi^* - \gamma^* \xi)/2} \, \hat D(\gamma + \xi)$, for the generic complex variables $\gamma$ and $\xi$. 
Applying the coherent states from the right and the optical Fock states from the left, we find 
\begin{align} \label{app:eq:photon:mum:intermediate}
\braket{\hat N_a(\tau)} &= e^{- |\alpha|^2} \sum_{n = 0}^\infty \sum_{m = 0}^\infty n \, \frac{|\alpha|^{2n}}{n!} e^{-  \tilde{\kappa}_{\rm{c}} \tau n}\,\bra{m} \otimes \bra{m} \left( e^{- i \,F_+n \, \hat B_+} \, e^{- i \,F_- n \, \hat B_-} \right)   \otimes \left( e^{i \,F_+ n \, \hat B_+}  \, e^{i \,F_-n \, \hat B_-} \,   \right) \nonumber \\
&\quad\times   \biggl[ 1  + \tilde{\kappa}_{\rm{c}} |\alpha|^2\int^\tau_0 \mathrm{d}\tau'  \, e^{- \tilde{\kappa}_{\rm{c}} \tau'} \, e^{-2 \,i \, A (\tau')\,n}\,\hat D(G(\tau'))   \,   \otimes  e^{2 \, i \,  A(\tau')\, n} \,   \hat D(G^*(\tau'))  \, \nonumber \\
&\quad \qquad + \tilde{\kappa}_{\rm{c}}^2 \, |\alpha|^4 \int^\tau_0 \mathrm{d}\tau' \, \int^{\tau'}_0 \mathrm{d}\tau'' \, e^{- \tilde{\kappa}_{\rm{c}}(\tau' + \tau'')} \, e^{-2 \, i \,  [A (\tau') + A(\tau'')]\,n }  \hat D(G(\tau') + G(\tau'')) \nonumber \\
&\quad\qquad \qquad \qquad \qquad \otimes  e^{2\, i \, [A(\tau') + A(\tau'')]\, n} \,   \, \hat D(G^*(\tau') + G^*(\tau''))  + \cdots   \biggr] \ket{\beta} \otimes \ket{\beta^*}+\cdots . 
\end{align}
We find that all phases inside the integrals cancel. It remains  to apply the displacement operators to the mechanical coherent state. We first note that the exponentials in the first line of~\eqref{app:eq:photon:mum:intermediate} can be written as displacement operators using the relations
\begin{align}
e^{- i \,  F_+ n  \,\hat B_+ } e^{- i \, F_-n  \,  \hat B_-} &= e^{  (F_- - i \,F_+ ) \, n \,\hat b^\dag - (F_- - i \,F_+)^* \,  n  \, \hat b} \, e^{-  i   \,  F_+ F_-\, n^2}  = \hat D\left(   n  \,G(\tau)\right) \, e^{-  i  \,  F_+ F_-\, n^2  },\nonumber \\
e^{ i \,  F_+n \, \hat B_+ } e^{ i \,F_- n \,  \hat B_-} &= e^{  (F_- + i \,F_+ ) \, n \,\hat b^\dag -  (F_- + i \,F_+)^* \,  n \,\hat b} \, e^{  i   \,  F_+ F_-\, n^2}  = \hat D( n\,G^*(\tau)) \, e^{ i   \,  F_+ F_-\, n^2}.
\end{align}
Then, applying everything to the mechanical states, we find  
\begin{align}
\braket{\hat N_a(\tau)} &= e^{- |\alpha|^2} \sum_{n = 0}^\infty \sum_{m = 0}^\infty n \, \frac{|\alpha|^{2n}}{n!} e^{-   \tilde{\kappa}_{\rm{c}} \tau \, n}\,\bra{m} \otimes \bra{m}  \biggl[ \ket{\beta + n \,G(\tau)}\ket{\beta^* + n \,G^*(\tau)}  \nonumber \\
&\, +  \tilde{\kappa}_{\rm{c}} |\alpha|^2 \int^\tau_0 \mathrm{d}\tau'  \, e^{-  \tilde{\kappa}_{\rm{c}} \tau'} \, \ket{\beta + G(\tau') + n\, G(\tau)}   \,   \otimes  \ket{\beta^* + G^*(\tau') + n\, G^*(\tau)} \,  \\
&\, +  \tilde{\kappa}_{\rm{c}}^2 \, |\alpha|^4 \int^\tau_0 \mathrm{d}\tau' \, \int^{\tau'}_0 \mathrm{d}\tau'' \, e^{- \tilde{\kappa}_{\rm{c}}(\tau' + \tau'')} \,    \ket{\beta + G(\tau') + G(\tau'') + n\, G(\tau)} \otimes   \, \ket{\beta^* + G^*(\tau') + G^*(\tau'') + n\, G^*(\tau)}  + \ldots   \biggr].  \nonumber
\end{align}
However, we now employ the  normalization condition:
\begin{equation}\label{normalization:condition}
\sum_{m = 0}^\infty \bra{m} \otimes \bra{m} \ket{\beta} \otimes \ket{\beta^*}=\sum_{m = 0}^\infty \braket{m |\beta}\braket{m |\beta^*} =  e^{- |\beta|^2} \sum_{m = 0}^\infty \frac{ |\beta|^{2m}}{m!} = 1, 
\end{equation}
which means that the terms inside the integral are all unity. Therefore, we are left with the expression
\begin{align}
\braket{\hat N_a(\tau)} &= e^{- |\alpha|^2} \sum_{n = 0}^\infty n \, \frac{|\alpha|^{2n}}{n!} e^{-   \tilde{\kappa}_{\rm{c}} \tau\, n}\,   \biggl[ 1  +  \tilde{\kappa}_{\rm{c}} \,  |\alpha|^2 \int^\tau_0 \mathrm{d}\tau'  \, e^{-  \tilde{\kappa}_{\rm{c}} \tau'} +  \tilde{\kappa}_{\rm{c}}^2 \, |\alpha|^4 \int^\tau_0 \mathrm{d}\tau' \, \int^{\tau'}_0 \mathrm{d}\tau'' \, e^{-  \tilde{\kappa}_{\rm{c}}(\tau' + \tau'')} + \cdots \biggr] .
\end{align}
This can be simplified further by collating all terms in the expansion and we obtain
\begin{equation}
\braket{\hat N_a(\tau)} = e^{- |\alpha|^2} \sum_{n = 0}^\infty n \, \frac{|\alpha|^{2n}}{n!} e^{-  \tilde{\kappa}_{\rm{c}} \tau \, n}\,   \mathrm{exp} \left[ \tilde{\kappa}_{\rm{c}} \, |\alpha|^2 \int^\tau_0 \mathrm{d} \tau' \, e^{- \tilde{\kappa}_{\rm{c}} \, \tau'} \right]. 
\end{equation}
Evaluating the integral and simplifying the expression, we find 
\begin{equation}
\braket{\hat N_a(\tau)}= |\alpha|^2 \, e^{- \tilde{\kappa}_{\rm{c}}  \, \tau}. 
\end{equation}
This expression is completely independent of the mechanical dynamics, which follows because the photon number operator commutes with the optomechanical Hamiltonian. 

\section{Derivation of the homodyne signal in Equation~\eqref{eq:homodyne:signal}} \label{app:homodyne}

In this appendix, we compute the expectation value of the annihilation operator $\hat a$ as a function of time $\tau$. 

\subsection{Coherent states}
We recall that, for vectorized states, the trace operation can be written as
\begin{equation} \label{app:eq:vectorized:trace}
\mathrm{Tr}\left[ \hat A^\dag \hat B\right] = \bbraket{A | B}. 
\end{equation}
For $\braket{\hat a(\tau)}$, we therefore have 
\begin{equation}
\braket{\hat a  (\tau)} = \mathrm{Tr}\left[ \hat a  \, \hat \varrho (\tau) \right] =  \bbra{\hat a^\dag  } \hat{\mathcal{S}}(\tau)\kket{\varrho_0}, 
\end{equation}
where $\kket{\varrho_0}$ is the vectorized initial density matrix. To find the vectorized state $\bbra{\hat a^\dag \otimes \mathds{1}}$ we expand the operator $\hat a^\dag$ in the Fock basis as follows:
\begin{equation}
\hat a^\dag = \hat a^\dag \sum_{n = 0}^\infty \ket {n}\bra {n} = \sum_{n = 0}^\infty \sqrt{n + 1} \ket{n+1} \bra{n}. 
\end{equation}
This allows us to write the vectorized operator that acts on both the optical and mechanical subsystems as
\begin{equation}
\bbra{\hat a^\dag  } = \sum_{n = 0}^\infty \sum_{m = 0}^\infty \sqrt{n+1} \bra{n+1}\bra{m} \otimes  \bra{n} \bra{m}. 
\end{equation}
Employing the initially vectorized state $\kket{\varrho_0 } = \ket{\alpha} \ket{\beta} \otimes \ket{\alpha^*} \ket{\beta^*}$ allows us to write the expectation value of $\hat a $ as
\begin{align}
\braket{\hat a(\tau)} &=  \sum_{n = 0}^\infty \sum_{m = 0}^\infty \sqrt{n+1} \bra{n+1}  \bra{m} \otimes \bra{n} \bra{m} \left( e^{- i \,\hat N_b\, \tau} \, e^{- i \,F_a \, \hat N_a^2} \, e^{- i \,F_+ \hat N_a \, \hat B_+} \, e^{- i F_- \hat N_a \, \hat B_-} \, e^{- \tilde{\kappa}_{\rm{c}} \tau  \hat N_a /2}\right)  \nonumber \\
&\qquad \qquad \qquad\qquad\qquad\qquad\qquad  \otimes \left( e^{ i \,\hat N_b\, \tau} \, e^{ i \,F_a\, \hat N_a^2} \, e^{ i \,F_+ \hat N_a \, \hat B_+} \, e^{ i \,F_- \hat N_a \, \hat B_-}e^{-  \tilde{\kappa}_{\rm{c}} \tau  \hat N_a /2}\right) \nonumber \\
&\quad \times \overleftarrow{\mathcal{T}} \mathrm{exp} \left[\tilde{\kappa}_{\rm{c}} \int^\tau_0 \mathrm{d}\tau'  \, e^{- \tilde{\kappa}_{\rm{c}} \tau'} \,e^{-2\,i\,A(\tau')\,\hat{N}_a}\,  \hat a\,\hat D(G(\tau'))    \otimes  e^{2 i A(\tau') \hat N_a} \, \hat a \,   \hat D(G^*(\tau')) \,  \right]\ket{\alpha} \ket{\beta} \otimes \ket{\alpha^*} \ket{\beta^*}. 
\end{align}
Applying the optical Fock states from the left and collecting some of the exponentials, we find 
\begin{align} \label{app:eq:homodyne:intermediate}
\braket{\hat a (\tau)} &=  \sum_{n = 0}^\infty \sum_{m = 0}^\infty \sqrt{n + 1} \, e^{- i \,F_a \, (2n+1)} \, e^{-  \tilde{\kappa}_{\rm{c}}  \, \tau  (2n + 1) /2} \bra{n+1} \bra{m} \otimes  \bra{n} \bra{m} \left(  e^{- i \,F_+ (n+1) \, \hat B_+} \, e^{- i \,F_- (n+1) \, \hat B_-}\right)   \otimes \left( e^{ i\, F_+ n \, \hat B_+} \, e^{ i \,F_- n \, \hat B_-}\right) \nonumber \\
&\quad \times \overleftarrow{\mathcal{T}} \mathrm{exp} \left[\tilde{\kappa}_{\rm{c}} \int^\tau_0 \mathrm{d}\tau'  \, e^{- \tilde{\kappa}_{\rm{c}} \tau'} \,e^{-2\,i\, A(\tau')\,\hat{N}_a}  \,  \hat a\,\hat D(G(\tau'))  \otimes  e^{2\, i \,A(\tau') \hat N_a} \, \hat a\,   \hat D(G^*(\tau'))  \,  \right]\ket{\alpha} \ket{\beta} \otimes \ket{\alpha^*} \ket{\beta^*}. 
\end{align}
The exponentials containing operators $\hat B_+$ and $\hat B_-$ can be combined into displacement operators as noted before. We have
\begin{align}
e^{- i \, (n + 1) \, F_+ \hat B_+ } e^{- i \, (n + 1) \, F_- \hat B_-} &= e^{ (n + 1) \, (F_- - i F_+ ) \hat b^\dag - (n + 1) \, (F_- - i F_+)^* \hat b} \, e^{-  i \, (n + 1)^2  \,  F_+ F_-}  = \hat D\left( (n + 1) G(\tau) \right) \, e^{-  i \, (n + 1)^2  \,  F_+ F_- },\nonumber \\
e^{ i \, n \, F_+ \hat B_+ } e^{ i \, n \, F_- \hat B_-} &= e^{ n \, (F_- + i F_+ ) \hat b^\dag -  n \, (F_- + i F_+)^* \hat b} \, e^{  i \, n^2  \,  F_+ F_-}  = \hat D(n \, G^*(\tau)) \, e^{ i \, n^2  \,  F_+ F_-}, 
\end{align}
where we note that the phase simplifies to $e^{- i (2n +1) F_+F_-}$. 
Using $A(\tau) = F_a + F_+ F_-$, we then write~\eqref{app:eq:homodyne:intermediate} as 
\begin{align}
\braket{\hat a(\tau)} &=\sum_{n = 0}^\infty \sum_{m = 0}^\infty \sqrt{n + 1} \, e^{- i A(\tau) \, (2n+1)} \, e^{-  \tilde{\kappa}_{\rm{c}} \tau  (2n + 1)/2 } \, \bra{n+1}  \bra{m} \otimes \bra{n} \bra{m} \hat D( (n+1)G(\tau)) \otimes \hat D(n\,G^*(\tau)) \nonumber \\
&\quad \times \overleftarrow{\mathcal{T}} \mathrm{exp} \left[\tilde{\kappa}_{\rm{c}} \int^\tau_0 \mathrm{d}\tau'  \, e^{- \tilde{\kappa}_{\rm{c}} \tau'} \,e^{-2\,i\,A(\tau')\,\hat{N}_a}\,  \hat a\,\hat D(G(\tau'))    \otimes  e^{2 \,i\, A(\tau') \hat N_a} \, \hat a\,   \hat D(G^*(\tau'))  \,  \right]\ket{\alpha} \ket{\beta} \otimes \ket{\alpha^*} \ket{\beta^*}. 
\end{align}
We now expand the integral again, as shown in Appendix~\ref{app:expanding:the:exponential}. Overlapping the expressions with the optical Fock states and simplifying, we find 
\begin{align} \label{app:eq:temporary:integral}
&\bra{n+1} \otimes \bra{n}\overleftarrow{\mathcal{T}} \mathrm{exp}\left[ \tilde{\kappa}_{\rm{c}} \int^\tau_0 \mathrm{d}\tau'  \, e^{- \tilde{\kappa}_{\rm{c}} \tau'} \, e^{-2 \,i \, A (\tau')\,\hat{N}_a} \,  \hat a\,\hat D(G(\tau'))   \otimes  e^{ 2 \,i \, A(\tau')\hat N_a} \, \hat a\,   \hat D(G^*(\tau'))  \, \right]\ket{\alpha}\otimes \ket{\alpha^*} \nonumber \\
&=  e^{- |\alpha|^2}  \frac{ \alpha^{(n+1)} \alpha^{*n}}{\sqrt{n! (n+1)!}} \biggl[ 1  + \tilde{\kappa}_{\rm{c}} \,  |\alpha|^2 \int^\tau_0 \mathrm{d}\tau'  \, e^{- \tilde{\kappa}_{\rm{c}} \tau'} \, e^{-2 \, i \, A (\tau')\, }\,\hat D(G(\tau'))    \otimes   \hat D(G^*(\tau')) \nonumber \\
&+ \tilde{\kappa}_{\rm{c}}^2 \, |\alpha|^4 \int^\tau_0 \mathrm{d}\tau' \, \int^{\tau'}_0 \mathrm{d}\tau'' \, e^{- \tilde{\kappa}_{\rm{c}}(\tau' + \tau'')} \, e^{-2\, i  \,[A (\tau') + A(\tau'')]} \,  \hat D(G(\tau')) \hat D(G(\tau''))  \otimes   \hat D(G^*(\tau'))\,\hat D(G^*(\tau''))  + \cdots  \biggr]. 
\end{align}
We then apply the Weyl operators under the integrals in~\eqref{app:eq:temporary:integral} to the coherent states of the mechanics $\ket{\beta}$. We find that the phases vanish because of the vectorization. For example, 
\begin{equation}
\hat D(G(\tau)) \otimes \hat D(G^*(\tau)) \ket{\beta}\ket{\beta^*} = e^{( G \beta^* - G^* \beta)/2} e^{(G^* \beta - \beta^* G)/2} \ket{\beta + G}\ket{\beta^* + G^*} = \ket{\beta + G}\ket{\beta^* + G^*}. 
\end{equation}
This allows us to write 
\begin{align}
\braket{\hat a(\tau)} &=\sum_{n = 0}^\infty \sum_{m = 0}^\infty \sqrt{n + 1} \, e^{- i A(\tau) \, (2n+1)} \, e^{- \tilde{\kappa}_{\rm{c}} \tau  (2n + 1)/2 } \,  e^{- |\alpha|^2}  \frac{ \alpha^{(n+1)} \alpha^{*n}}{\sqrt{n! (n+1)!}} \bra{m}\otimes \bra{m} \hat D( (n+1)G(\tau)) \otimes \hat D(n\,G^*(\tau)) \nonumber \\
&\quad \times  \biggl[ \ket{\beta }\ket{\beta^* }  + \tilde{\kappa}_{\rm{c}} \,  |\alpha|^2 \int^\tau_0 \mathrm{d}\tau'  \, e^{- \tilde{\kappa}_{\rm{c}} \tau'} \, e^{-2 \, i \, A (\tau')\, }\,  \ket{\beta + G(\tau') }\ket{\beta^* + G^*(\tau') } \nonumber \\
&\qquad+ \tilde{\kappa}_{\rm{c}}^2 \, |\alpha|^4 \int^\tau_0 \mathrm{d}\tau' \, \int^{\tau'}_0 \mathrm{d}\tau'' \, e^{- \tilde{\kappa}_{\rm{c}}(\tau' + \tau'')} \, e^{-2\, i  \,[A (\tau') + A(\tau'')]} \,   \ket{\beta + G(\tau') + G(\tau'') + }\ket{\beta^* + G^*(\tau') + G^*(\tau'')}  + \cdots  \biggr]. 
\end{align}
Next, we apply the operators  $\hat D( (n+1)G(\tau)) \otimes \hat D(n\,G^*(\tau))$ to each expanded integral term. Here, we find that only part of the phases cancel. Starting with the zeroth-order contribution $\ket{\beta}\ket{\beta^*}$, we find 
\begin{align}
\hat D( (n+1)G(\tau)) \otimes \hat D(n\,G^*(\tau)) \ket{\beta}\ket{\beta^*} &= e^{(n+1)(G\beta^* - G^* \beta)/2} e^{n (G^*\beta - G \beta^*)/2} \ket{\beta + G(\tau)} \ket{\beta^* + G^*(\tau)} \nonumber \\
&=  e^{(G\beta^* - G^* \beta)/2} \ket{\beta + G(\tau)} \ket{\beta^* + G^*(\tau)}. 
\end{align}
The same can be done for any of the already displaced coherent states, such as $\ket{\beta + G(\tau')}$. Here, we must be careful to keep those terms that depend on $\tau'$ under the integral. However, since any displacement is a linear addition to $\beta$, we can write these terms as follows, taking the first-order contribution as an example:
\begin{equation}
 e^{\{G(\tau) [\beta^*  + G^*(\tau')] - G^*(\tau) [\beta + G(\tau')]\}/2}= e^{[G(\tau) \beta^* - G^*(\tau) \beta]/2} e^{[G(\tau) G^*(\tau') - G^*(\tau) G(\tau')]/2}. 
\end{equation}
That is, we can always divide these phases into one expression that depends on $\tau$ and $\beta$, and one that depends on $\tau$ and $\tau'$ (and any further $\tau''$ or $\tau'''$ etc.).

This allows us to write $\braket{\hat a (\tau)}$ as 
\begin{align}
\braket{\hat a(\tau)} &=\sum_{n = 0}^\infty \sum_{m = 0}^\infty \sqrt{n + 1} \, e^{- i A(\tau) \, (2n+1)} \, e^{- \tilde{\kappa}_{\rm{c}} \tau  (2n + 1)/2 } \,   e^{[G(\tau) \beta^* - G^*(\tau) \beta]/2} \, e^{- |\alpha|^2}  \frac{ \alpha^{(n+1)} \alpha^{*n}}{\sqrt{n! (n+1)!}}  \nonumber \\
&\bra{m}\otimes \bra{m}   \biggl[ \ket{\beta + (n + 1) G(\tau)}\ket{\beta^* + n \, G^*(\tau)}  \nonumber \\
&\qquad + \tilde{\kappa}_{\rm{c}} \,  |\alpha|^2 \int^\tau_0 \mathrm{d}\tau'  \, e^{- \tilde{\kappa}_{\rm{c}} \tau'} \, e^{-2 \, i \, A (\tau')\, }\, e^{[G(\tau) G^*(\tau') - G^*(\tau) G(\tau')]/2} \, \ket{\beta + G(\tau') + (n +1) G(\tau)}\ket{\beta^* + G^*(\tau') + n \, G^*(\tau)} \nonumber \\
&\qquad+ \tilde{\kappa}_{\rm{c}}^2 \, |\alpha|^4 \int^\tau_0 \mathrm{d}\tau' \, \int^{\tau'}_0 \mathrm{d}\tau'' \, e^{- \tilde{\kappa}_{\rm{c}}(\tau' + \tau'')} \, e^{-2\, i  \,[A (\tau') + A(\tau'')]} \, e^{\{G(\tau) [G^*(\tau') + G^*(\tau'') ] - G^*(\tau) [G(\tau') + G(\tau'')]\}/2} \,  \nonumber \\
&\qquad \qquad \qquad \times \ket{\beta + G(\tau') + G(\tau'') + (n +1) G(\tau)}\ket{\beta^* + G^*(\tau') + G^*(\tau'') + n \, G^*(\tau)}  + \cdots  \biggr]. 
\end{align}
Next, we must compute the overlap between the Fock states and the coherent states. We find the general expression for the generic complex variable $\gamma$:
\begin{align}
\sum_{m=0}^\infty \braket{m|\gamma +(n+1) G}\braket{m| \gamma^* +n\, G^*  }  &= e^{-|G|^2 /2} \, e^{[G \gamma^* - G^* \gamma]/2 }. 
\end{align}
Then, because each additional term from the expansions enters linearly since $\gamma = \beta + G(\tau') + G(\tau'') + \ldots$ into the exponentials, they can be collated as increasing orders of the expanded Neumann series. Note that this adds another factor of $e^{[G(\tau) \beta^* - G^*(\tau) \beta]/2}$ in the expression. 

Collecting all terms, it allows us to write the entire expression as
\begin{align}
\braket{\hat a(\tau)} &=e^{- |\alpha|^2}  \sum_{n = 0}^\infty \sqrt{n + 1}  \frac{ \alpha^{(n+1)}\alpha^{*n}}{\sqrt{n! (n+1)!}}  e^{-|G(\tau)|^2/2}\, e^{- i \, A(\tau) \, (2n+1)} \, e^{-  \tilde{\kappa}_{\rm{c}} \tau (2n + 1)/2 } \, e^{G(\tau) \beta^* - G^*(\tau) \beta}  \nonumber \\
&\quad \times  \mathrm{exp} \left[\tilde{\kappa}_{\rm{c}} \, |\alpha|^2 \int^\tau_0 \mathrm{d}\tau'  \, e^{- \tilde{\kappa}_{\rm{c}} \tau'} \,e^{-2\,i\,A(\tau')}  \, e^{[G(\tau) G^{*}(\tau') - G^*(\tau) G(\tau')]} \right]. 
\end{align}
This expression can be further simplified by evaluating the sum over $n$. We find 
\begin{equation}
e^{- |\alpha|^2} \sum_{n = 0}^\infty \frac{ \alpha^{n+1}\alpha^{*n} }{n!} e^{-\tilde{\kappa}_{\rm{c}} \tau \, (2 n + 1)/2} e^{ - i \, A(\tau) \,  (2n+1)} = \alpha \,  e^{|\alpha|^2 \left(e^{ -\tilde{\kappa}_{\rm{c}} \tau } e^{ -2\, i \, A(\tau)}-1\right)} e^{-\tilde{\kappa}_{\rm{c}} \tau/2} \, e^{- i \, A(\tau)}. 
\end{equation}
We then write the full expression as 
\begin{align} \label{app:eq:homodyne:signal}
\braket{\hat a(\tau)} &=\alpha \, e^{|\alpha|^2 \left(e^{ - 2\, i \, A(\tau)} \, e^{- \tilde{\kappa}_{\rm{c}}  \tau}-1\right) } e^{-|G(\tau)|^2/2}\,  e^{- i\, A(\tau)}  \, e^{ - \tilde{\kappa}_{\rm{c}} \tau/2} \, e^{G(\tau) \beta^* - G^*(\tau) \beta}  \, 	  \mathrm{exp} \left[\tilde{\kappa}_{\rm{c}} \, |\alpha|^2 \int^\tau_0 \mathrm{d}\tau'  \, e^{- \tilde{\kappa}_{\rm{c}} \tau'} \,e^{-2\,i\,A(\tau')}  \, e^{i \, B(\tau', \tau) } \right]. 
\end{align}
where we have defined $B(\tau', \tau) =  2 \, \mathrm{Im}[G(\tau) G^*(\tau')]$. 
The expression of $\braket{\hat a(\tau)}$ for $\tilde{\kappa}_{\rm{c}}= 0$ has been obtained in the literature~\cite{qvarfort2019enhanced} and it coincides with our expression of~\eqref{app:eq:homodyne:signal} for $\tilde{\kappa}_{\rm{c}}= 0$, that is, 
\begin{align}
\braket{\hat a (\tau)} &=  \, \alpha \,  e^{|\alpha|^2 \left(e^{ - 2\, i \,A(\tau) }  -1\right) } \, e^{-|G(\tau)|^2 /2}\, e^{- i \, A(\tau)} \, e^{G \beta^* - G^* \beta}. 
\end{align}

\subsection{Extension to thermal mechanical states} \label{app:quadratures}

As discussed in the main text, thermal mechanical states are a much better representation of realistic conditions. The expression~\eqref{app:eq:homodyne:signal} can be generalized for thermal states by a weighted integration over $\beta$
\begin{align}
\braket{\hat a(\tau)}_{\mathrm{th}} &=\alpha \, e^{|\alpha|^2 \left(e^{ - 2\, i \, A(\tau)} \, e^{- \tilde{\kappa}_{\rm{c}}  \tau}-1\right) } e^{-|G(\tau)|^2/2}\,  e^{- i\, A(\tau)}  \, e^{ -  \tilde{\kappa}_{\rm{c}} \tau/2}   \, 	  \mathrm{exp} \left[\tilde{\kappa}_{\rm{c}} \, |\alpha|^2 \int^\tau_0 \mathrm{d}\tau'  \, e^{- \tilde{\kappa}_{\rm{c}} \tau'} \,e^{-2\,i\,A(\tau')}  \, e^{i \, B(\tau', \tau) } \right] \, \nonumber \\
&\times \frac{1}{\bar{n}\pi} \int^\infty_{-\infty} \mathrm{d}^2 \beta \, e^{- |\beta|^2/\bar{n}} \, e^{G(\tau) \beta^* - G^*(\tau) \beta}, 
\end{align}
where $\beta\in\mathbb{C}$ and where $\bar{n}$ is the average phonon number in the system. 

Expanding $\beta$ in terms of a real and complex part, such that $\beta = x + i y$, we can write the integral as
\begin{align}
\frac{1}{\bar{n}\pi} \int^\infty_{-\infty} \mathrm{d}^2 \beta \, e^{- |\beta|^2/\bar{n}} \, e^{G(\tau) \beta^* - G^*(\tau) \beta} &=  \frac{1}{\bar{n}\pi} \int^\infty_{-\infty} \mathrm{d}x \mathrm{d}y \, e^{- (x^2 + y^2)/\bar{n}} \, e^{G(\tau) (x - i y) - G^*(\tau) (x + i y)} =  e^{-|G|^2 n}.
\end{align}
The expectation value becomes 
\begin{align} \label{app:eq:a:thermal}
\braket{\hat a(\tau)}_{\mathrm{th}} &= \alpha \, e^{|\alpha|^2 \left(e^{ - 2\, i \, A(\tau)} \, e^{- \tilde{\kappa}_{\rm{c}}  \tau}-1\right) } e^{-|G(\tau)|^2(1 + 2 \bar{n})/2}\,  e^{- i\, A(\tau)}  \, e^{ -  \tilde{\kappa}_{\rm{c}} \tau/2}   \, 	  \mathrm{exp} \left[\tilde{\kappa}_{\rm{c}} \, |\alpha|^2 \int^\tau_0 \mathrm{d}\tau'  \, e^{- \tilde{\kappa}_{\rm{c}} \tau'} \,e^{-2\,i\,A(\tau')}  \, e^{i \, B(\tau', \tau) } \right] .
\end{align}

\subsection{Behaviour of the quadratures in the long-time limit} \label{app:quadratures}

We wish to determine what happens to the quadratures for nonzero $\tilde{\kappa}_{\rm{c}}$. They are defined as $\braket{\hat X_{\rm{c}}(\tau)} = \sqrt{2} \, \mathrm{Re} \braket{\hat a(\tau)}$ and $\braket{\hat P_{\rm{c}}(\tau)} = \sqrt{2} \, \mathrm{Im} \braket{\hat a (\tau)} $.  To examine these expressions, which both depend on $\braket{\hat a(\tau)}$, we start by examining $|\braket{\hat a(\tau)}|^2$, which we can bound effectively. The modulus becomes 
\begin{align}
|\braket{\hat a(\tau)}|^2 &= |\alpha|^2 \, e^{|\alpha|^2 \left(2 \cos[ 2\,A(\tau)] \, e^{- \tilde{\kappa}_{\rm{c}}  \tau}-2\right) } 
\, e^{-|G(\tau)|^2} \, e^{ - \tilde{\kappa}_{\rm{c}} \tau} \,   \mathrm{exp} \left[2 \, \tilde{\kappa}_{\rm{c}} \, |\alpha|^2 \int^\tau_0 \mathrm{d}\tau'  \, e^{- \tilde{\kappa}_{\rm{c}} \tau'} \,\cos\left[2\,A(\tau') -  B(\tau', \tau) \right] \right] .
\end{align}
The integral is bounded by the maximum value of its argument, in the sense that
\begin{align}
\int \mathrm{d}x f_{\rm{min}}(x) \leq \int \mathrm{d}x \, f(x) \leq \int \mathrm{d}x \, f_{\rm{max}}(x). 
\end{align}
We note that the maximum value of $\cos(x)$ is 1. Inserting this, we find that 
\begin{align}
|\braket{\hat a(\tau)}|^2 &\leq |\alpha|^2 \, e^{|\alpha|^2 \left(2 \cos[ 2\,A(\tau)] \, e^{- \tilde{\kappa}_{\rm{c}}  \tau}-2\right) } 
\, e^{-|G(\tau)|^2} \, e^{ - \tilde{\kappa}_{\rm{c}} \tau} \,   \mathrm{exp} \left[2 \, \tilde{\kappa}_{\rm{c}} \, |\alpha|^2 \int^\tau_0 \mathrm{d}\tau'  \, e^{- \tilde{\kappa}_{\rm{c}} \tau'} \right] .
\end{align}
The integral evaluates to 
\begin{equation}
\int^\tau_0 \mathrm{d}\tau'  \, e^{- \tilde{\kappa}_{\rm{c}} \tau'} = \frac{1}{\tilde{\kappa}_{\rm{c}}} \left( 1 - e^{- \tau \tilde{\kappa}_{\rm{c}}} \right), 
\end{equation}
which means that we are left with 
\begin{align}
|\braket{\hat a(\tau)}|^2 &\leq |\alpha|^2 \, e^{2 \, |\alpha|^2  \cos[ 2\,A(\tau)] \, e^{- \tilde{\kappa}_{\rm{c}}  \tau} } 
 e^{-|G(\tau)|^2} \, e^{ - \tilde{\kappa}_{\rm{c}} \tau} \, e^{-2 \, |\alpha|^2 \, e^{- \tau \, \tilde{\kappa}_{\rm{c}}}}. 
\end{align}
The presence of the term $e^{- \tau \, \tilde{\kappa}_{\rm{c}}}$ means that for $\tilde{\kappa}_{\rm{c}} \neq 0$, $|\braket{\hat a (\tau) }|^2 \rightarrow 0$ as $\tau \rightarrow \infty$. Therefore, we conclude that the two quadratures tend to zero in the long-time limit.

\section{Derivation of the fidelity for preparing noisy optical cat states} \label{app:fidelity}
In the vectorized language, the density matrix of the two coherent states is given by 
\begin{equation}
\kket{\varrho_0} = \ket{\alpha}\ket{\beta}\otimes \ket{\alpha^*} \ket{\beta^*}. 
\end{equation}
Applying $\hat{\mathcal{S}}(\tau)$ to this initial state, we find that the evolved noisy state is given by 
\begin{align}\label{app:eq:full:noisy:state}
\kket{\varrho(\tau)} &=  \left( e^{- i \, \hat N_b\, \tau} \, e^{- i\,  F_a \, \hat N_a^2} \, e^{- i \, F_+ \hat N_a \, \hat B_+} \, e^{- i \, F_- \hat N_a \, \hat B_-}e^{- \tilde{\kappa}_{\rm{c}} \tau \hat N_a/2 }\right)   \otimes \left( e^{i \, \hat N_b\, \tau} \, e^{ i\,  F_a \, \hat N_a^2} \, e^{ i \, F_+ \hat N_a \, \hat B_+} \, e^{ i \, F_- \hat N_a \, \hat B_-}e^{-  \tilde{\kappa}_{\rm{c}} \tau \hat N_a/2 }\right)    \\
&\quad \times \overleftarrow{\mathcal{T}} \mathrm{exp} \left[\tilde{\kappa}_{\rm{c}} \int^\tau_0 \mathrm{d}\tau'  \, e^{- \tilde{\kappa}_{\rm{c}} \tau'} \,e^{-2\,i\,A(\tau')\,\hat{N}_a}\,\hat D(G(\tau))  \,  \hat a \otimes  e^{2\, i\, A(\tau') \hat N_a} \,   \hat D(G^*(\tau')) \, \hat a \,  \right] \ket{\alpha}\ket{\beta}\otimes \ket{\alpha^*} \ket{\beta^*} \nonumber , 
\end{align}
where we recall that $A(\tau') = F_a + F_+ F_-$. 
For a constant optomechanical coupling, the $F$ coefficients are given in~\eqref{eq:F:coefficients}. At  $\tau = 2\pi$ we find $F_+= F_-  = 0$, and also $G(2\pi) = 0$. The state in~\eqref{app:eq:full:noisy:state} can be simplified to
\begin{align} 
\kket{\varrho(2\pi)} &=  \left( e^{- i \,F_a \, \hat N_a^2} \,e^{- \pi \tilde{\kappa}_{\rm{c}}  \hat N_a }\right)   \otimes \left( e^{ i \,F_a \, \hat N_a^2} \,e^{- \pi \tilde{\kappa}_{\rm{c}}  \hat N_a }\right)    \\
&\quad \times \overleftarrow{\mathcal{T}} \mathrm{exp} \left[\tilde{\kappa}_{\rm{c}} \int^{2\pi}_0 \mathrm{d}\tau'  \, e^{- \tilde{\kappa}_{\rm{c}} \tau'} \,e^{-2\,i\,A(\tau')\,\hat{N}_a}\,\hat D(G(\tau))  \,  \hat a \otimes  e^{2\, i \,A(\tau') \hat N_a} \,   \hat D_(G^*(\tau')) \, \hat a \,  \right] \ket{\alpha}\ket{\beta}\otimes \ket{\alpha^*} \ket{\beta^*}. \nonumber
\end{align}
We then trace out the mechanical state with $\hat \varrho_{\rm{c}}(\tau) = \mathrm{Tr}_{\rm{m}} \left[ \hat \varrho(\tau) \right] $. The tracing operation in the vectorized language involves taking the overlap with the identity, which we resolve in terms of the mechanical Fock states as $\sum_{m = 0}^\infty \bra{m}\bra{m}$. The cavity state $\kket{\varrho(2\pi)}_{\rm{c}}$ at $\tau=2\,\pi$ is then given by 
\begin{align}
\kket{\varrho(2\pi)}_{\rm{c}} &= \sum_{m = 0}^\infty \bra{m}\otimes\bra{m}\kket{\varrho(2\pi)} \nonumber \\
& = \sum_{m = 0}^\infty \bra{m} \otimes \bra{m} \left( e^{- i F_a \, \hat N_a^2} \,  e^{-\pi \tilde{\kappa}_{\rm{c}}  \hat N_a }\right)   \otimes \left( e^{ i F_a \, \hat N_a^2} \,   e^{- 2\pi \tilde{\kappa}_{\rm{c}}  }\right)    \\
&\quad \times \overleftarrow{\mathcal{T}} \mathrm{exp} \left[\tilde{\kappa}_{\rm{c}} \int^{2\pi}_0 \mathrm{d}\tau'  \, e^{- \tilde{\kappa}_{\rm{c}} \tau'} \,e^{-2\,i\,A(\tau')\,\hat{N}_a}\,\hat D(G(\tau'))  \,  \hat a \otimes  e^{2\, i \,A(\tau') \hat N_a} \,   \hat D(G^*(\tau')) \, \hat a \,  \right] \ket{\alpha}\ket{\beta}\otimes \ket{\alpha^*} \ket{\beta^*}. \nonumber
\end{align}
We then expand the exponential according to the expression in~\eqref{app:eq:simplified:exponential}. The traced-out cavity state therefore becomes
\begin{align}
\kket{\varrho(2\pi)}_{\rm{c}} &= \sum_{m = 0}^\infty \bra{m} \otimes \bra{m} \left( e^{- i F_a \, \hat N_a^2} \,  e^{-\pi \tilde{\kappa}_{\rm{c}}  \hat N_a }\right)   \otimes \left( e^{ i F_a \, \hat N_a^2} \,   e^{- \pi \tilde{\kappa}_{\rm{c}}  \hat N_a }\right)  \nonumber  \\
&\times \biggl[  1  + \tilde{\kappa}_{\rm{c}} \int^\tau_0 \mathrm{d}\tau'  \, e^{- \tilde{\kappa}_{\rm{c}} \tau'} \, e^{-2\,i \,A (\tau')\,\hat{N}_a}\,\hat D(G(\tau'))   \,  \hat a \otimes  e^{2\, i \,A(\tau')\hat N_a} \,   \hat D(G^*(\tau')) \, \hat a \,  \\
&+ \tilde{\kappa}_{\rm{c}}^2 \int^\tau_0 \mathrm{d}\tau' \, \int^{\tau'}_0 \mathrm{d}\tau'' \, e^{- \tilde{\kappa}_{\rm{c}}(\tau' + \tau'')} \, e^{-2\,i \,[A (\tau') + A(\tau'')]\,\hat{N}_a}\,\hat a ^2 \,  \hat D(G(\tau')) \, \hat D(G(\tau''))  \nonumber \\
&\qquad \qquad \qquad\qquad  \otimes  e^{2\, i\, [A(\tau') + A(\tau'')]\hat N_a} \,  \hat a^2 \, \hat D(G^*(\tau'))\,\hat D(G^*(\tau''))  + \cdots \, \biggr] \ket{\alpha}\ket{\beta}\otimes \ket{\alpha^*} \ket{\beta^*}. \nonumber
\end{align}
However, we note that for each order of the expansion, the overlap between the mechanical Fock states and the coherent states satisfies the normalization condition~\eqref{normalization:condition}, which implies that we are left with the following expression for the traced-out cavity state:
\begin{align}
\kket{\varrho(2\pi)}_{\rm{c}} &=  \left( e^{- i \,F_a \, \hat N_a^2} \,e^{- \pi \tilde{\kappa}_{\rm{c}}  \hat N_a }\right)   \otimes \left( e^{ i \,F_a \, \hat N_a^2} \,e^{- \pi \tilde{\kappa}_{\rm{c}}  \hat N_a }\right) \overleftarrow{\mathcal{T}} \mathrm{exp} \left[\tilde{\kappa}_{\rm{c}} \int^{2\pi}_0 \mathrm{d}\tau'  \, e^{- \tilde{\kappa}_{\rm{c}} \tau'} \,e^{-2\,i\,A(\tau')\,\hat{N}_a}\, \hat a \otimes  e^{2\, i \,A(\tau') \hat N_a} \, \hat a \,  \right] \ket{\alpha} \otimes \ket{\alpha^*}. 
\end{align}
The fidelity $\mathcal{F}$ is then given by the overlap between the ideal cat state and the mixed state as $\mathcal{F} = \bra{\Psi(2\pi)} \hat \varrho (\tau) \ket{\Psi(2\pi)}$. In the vectorized language, the ideal cat state is given by 
\begin{align}
\kket{\Psi(2\pi)}_{\rm{c}} = \ket{\Psi(2\pi)}_{\rm{c}} \otimes \ket{\Psi^*(2\pi)}_{\rm{c}} =  e^{- |\alpha|^2} \sum_{n = 0}^\infty \sum_{n' = 0}^\infty \frac{\alpha^n \alpha^{*n'}}{\sqrt{n!n'!}} e^{- i F_a \, (n^2 - n^{\prime 2})} \ket{n}_{\rm{c}} \otimes \ket{n'}_{\rm{c}} .
\end{align}
The overlap becomes
\begin{align}
\mathcal{F}(2\,\pi) &= \bbraket{\Psi^\dag(2\pi) |\varrho(2\pi)} \nonumber \\
&= e^{- |\alpha|^2} \sum_{n = 0}^\infty \sum_{n' = 0}^\infty \frac{\alpha^{*n} \alpha^{n'}}{\sqrt{n! n'!}} e^{ i \,F_a (n^2 -n^{\prime 2})} \bra{n} \otimes \bra{n'}  \left( e^{- i F_a \, \hat N_a^2} \, e^{- \pi  \tilde{\kappa}_{\rm{c}} \hat N_a }\right)   \otimes \left( e^{ i \,F_a \, \hat N_a^2} \,  e^{- \pi \tilde{\kappa}_{\rm{c}} \hat N_a }\right)   \nonumber \\
&\quad \times \overleftarrow{\mathcal{T}} \mathrm{exp} \left[\tilde{\kappa}_{\rm{c}} \int^{2\pi}_0 \mathrm{d}\tau'  \, e^{- \tilde{\kappa}_{\rm{c}} \tau'} \,e^{-2\,i\,A(\tau')\,\hat{N}_a}\, \,  \hat a \otimes  e^{2\, i \,A(\tau') \hat N_a} \,  \hat a \,  \right]  \ket{\alpha} \otimes \ket{\alpha^*} . 
\end{align}
We apply the Fock states from the left to find that some of the phases cancel. We are left with
\begin{align}
\mathcal{F}(2\,\pi) &=  e^{- |\alpha|^2} \sum_{n = 0}^\infty \sum_{n' = 0}^\infty \frac{\alpha^{*n} \alpha^{n'}}{\sqrt{n! n'!}} \, e^{- \pi  \tilde{\kappa}_{\rm{c}} (n+n')} \,\bra{n} \otimes \bra{n'}   \overleftarrow{\mathcal{T}} \mathrm{exp} \left[\tilde{\kappa}_{\rm{c}} \int^{2\pi}_0 \mathrm{d}\tau'  \, e^{- \tilde{\kappa}_{\rm{c}} \tau'} \,e^{-2\,i\,A(\tau')\,\hat N_a}\, \,  \hat a \otimes  e^{2 \,i \, A(\tau') \, \hat N_a} \, \hat a \,  \right]  \ket{\alpha} \otimes \ket{\alpha^*}. 
\end{align}
We now attempt to simplify the integral. By using again the expansion of the integral in~\eqref{app:eq:simplified:exponential}, and applying the coherent states as done in Appendixes~\ref{app:photon:number} and~\ref{app:homodyne}, we find that the fidelity is given by 
\begin{align} \label{app:eq:fidelity}
\mathcal{F}(2\,\pi) &=  e^{- |\alpha|^2} \sum_{n = 0}^\infty \sum_{n' = 0}^\infty \frac{\alpha^{*n} \alpha^{n'}}{\sqrt{n! n'!}} \, e^{- \pi \tilde{\kappa}_{\rm{c}} (n+n')} \,\bra{n} \otimes \bra{n'} \mathrm{exp} \left[\tilde{\kappa}_{\rm{c}} |\alpha|^2 \int^{2\pi}_0 \mathrm{d}\tau' e^{- \tilde{\kappa}_{\rm{c}} \tau'} \, e^{- i A(\tau') (n-n')}
\right] \ket{\alpha} \otimes \ket{\alpha^*} . 
\end{align}
We note that setting $\tilde{\kappa}_{\rm{c}} = 0$ causes the exponential to vanish and the sums can be evaluated to recover $\mathcal{F}(2\,\pi) = 1$, as expected.

\subsection{Simplifying the fidelity}
To further simplify the expression~\eqref{app:eq:fidelity}, we divide the sum into three parts: a sum over diagonal elements with $n = n'$ and two sums where $n>n'$ and $n<n'$. The two second sums can be written as a single sum by renaming the index. The expression then reads
\begin{align}
\mathcal{F}(2\,\pi) &=  e^{- 2 \,|\alpha|^2} \sum_{n = 0}^\infty \frac{|\alpha|^{4n} }{(n!)^2}  e^{-2 \pi \tilde{\kappa}_{\rm{c}} n} \mathrm{exp} \left[\tilde{\kappa}_{\rm{c}} |\alpha|^2 \int^{2\pi}_0 \mathrm{d}\tau' e^{- \tilde{\kappa}_{\rm{c}} \tau'} \right]
\nonumber \\
& \quad+2 \, \mathrm{Re}\left\{ e^{- 2\,|\alpha|^2} \sum_{n = 1}^\infty \sum_{ n' = 0}^\infty  \frac{|\alpha|^{2( n + n')}}{n! n'! } e^{- \tilde{\kappa}_c \pi ( n + n')} \mathrm{exp} \left[\tilde{\kappa}_c |\alpha|^2 \int^{2\pi}_0 \mathrm{d}\tau' e^{- \tilde{\kappa}_{\rm{c}}  \tau'} \, e^{- 2iA(\tau') (n- n')}\right]\right\}. 
\end{align}
The integral in the first term evaluates to 
\begin{align} \label{app:eq:evaluated:integral}
 \int^\tau_ 0 \mathrm{d}\tau' e^{- \tilde{\kappa}_{\rm{c}} \tau'}  = \frac{1}{\tilde{\kappa}_{\rm{c}}} \left(1-e^{-\tilde{\kappa}_{\rm{c}} \tau}\right). 
\end{align}
We then define the index $ k = n - n' $, which runs from 1 to infinity due to the fact that we assumed that $n>n'$ at all times. We find 
\begin{align}
\mathcal{F}(2\,\pi) &=  e^{- 2|\alpha|^2} \sum_{n=0}^\infty \frac{|\alpha|^{4n} }{(n!)^2}  e^{-2 \pi \tilde{\kappa}_{\rm{c}} n} e^{|\alpha|^2 \left(1-e^{-2 \pi \tilde{\kappa}_{\rm{c}}}\right)}
\nonumber \\
&+2 \, \mathrm{Re}\biggl\{ e^{- 2|\alpha|^2} \sum_{n' = 0}^\infty \sum_{k = 1}^\infty \frac{|\alpha|^{4 n' + 2k}}{n'! (n' + k)!} e^{- \tilde{\kappa}_{\rm{c}} \pi ( 2n' + k)} \mathrm{exp} \left[\tilde{\kappa}_{\rm{c}} |\alpha|^2 \int^{2\pi}_0 \mathrm{d}\tau' e^{- \tilde{\kappa}_{\rm{c}}  \tau'} \, e^{- 2iA(\tau') k}\right]. 
\end{align}
We can now evaluate the diagonal sum and the second sum over $n'$. We find 
\begin{align}
\sum_{n = 0}^\infty \frac{|\alpha|^{4n}}{(n!)^2} &= e^{-2 \pi  \tilde{\kappa}_{\rm{c}} } I_0\left(2 \, |\alpha|^2\right) ,\nonumber \\
\sum_{n' = 0}^\infty \frac{|\alpha|^{4 n' + 2 k}}{n'! (n' + k)!} e^{- 2\pi \tilde{\kappa}_{\rm{c}} n'} &= |\alpha|^{- 2 k } \, e^{\pi \tilde{\kappa}_{\rm{c}} k} I_k \left( 2\, |\alpha|^2 \, e^{- \pi \tilde{\kappa}_{\rm{c}}} \right), 
\end{align}
where $I_k(x)$ is the Bessel function of order $k$. The fidelity can then be written as
\begin{align} \label{app:eq:fidelity:Bessel:functions}
\mathcal{F}(2\,\pi) 
&= e^{- 2|\alpha|^2} I_0(2 |\alpha|^2 e^{- \tilde{\kappa}_{\rm{c}}  \pi} ) e^{|\alpha|^2 \left(1-e^{- 2 \pi \tilde{\kappa}_{\rm{c}} }\right)}+ 2e^{- 2|\alpha|^2 } \sum_{k = 1}^\infty I_k (2 |\alpha|^2 e^{-\tilde{\kappa}_{\rm{c}} \pi} ) \mathrm{Re} \left\{  \mathrm{exp} \left[ \tilde{\kappa}_{\rm{c}}  |\alpha|^2 \int^{2\pi}_0 \mathrm{d} \tau' \, e^{- \tilde{\kappa}_{\rm{c}}  \tau'} e^{- 2 iA(\tau') k } \right] \right\}. 
\end{align}
Focusing on the second term in~\eqref{app:eq:fidelity:Bessel:functions}, which we call $\mathcal{F}_2$ we Taylor expand the exponential to find
\begin{align}
\mathcal{F}_2 &= 2\, e^{- 2|\alpha|^2 } \sum_{k = 1}^\infty I_k (2 \, |\alpha|^2 e^{-\tilde{\kappa}_{\rm{c}} \pi} ) \mathrm{Re} \left\{  \mathrm{exp} \left[ \tilde{\kappa}_{\rm{c}}  |\alpha|^2 \int^{2\pi}_0 \mathrm{d} \tau' \, e^{- \tilde{\kappa}_{\rm{c}}  \tau'} e^{- 2 iA(\tau') k } \right] \right\}  \nonumber\\
&= 2\, e^{- 2|\alpha|^2} \sum_{k = 1}^\infty I_k(2 \, |\alpha|^2 e^{- \pi \tilde{\kappa}_{\rm{c}}} ) \mathrm{Re} \biggl\{1 + \sum_{q = 1}^\infty \frac{( \tilde{\kappa}_{\rm{c}} |\alpha|^2)^q}{q!} \int^{2\pi}_0 \mathrm{d}\tau^{(1)} \cdots \mathrm{d}\tau^{(q)} e^{- \tilde{\kappa}_{\rm{c}} \sum_{p = 1}^q \tau^{(p)} } \, e^{2 i k \sum_{p = 1}^q A(\tau^{(p)}) } \biggr\} \nonumber \\
&= 2 \, e^{- 2|\alpha|^2} \sum_{k = 1}^\infty  \left[ I_k ( 2 \, |\alpha|^2 e^{ - \pi \tilde{\kappa}_{\rm{c}}} ) + \sum_{q = 1}^\infty \frac{(  \tilde{\kappa}_{\rm{c}}|\alpha|^2)^q}{q!} \int^{2\pi}_0 \mathrm{d}\tau^{(1)} \cdots \tau^{(q)} \, e^{- \tilde{\kappa}_{\rm{c}} \sum_{p = 1}^\infty \tau^{(p)} } I_k (2 \, |\alpha|^2 e^{- \pi \tilde{\kappa}_{\rm{c}}} ) \cos(2 k \sum_{p = 1}^q A( \tau^{(p)}) ) \right]. 
\end{align}
We then use the fundamental Jacobi-Anger expansion for the Bessel functions, which reads
\begin{align}
e^{z \cos \theta} = I_0(z) + 2 \sum_{n = 1}^\infty I_n(z) \, \cos(n \, \theta). 
\end{align}
This allows us to rewrite the sums over $k$ in terms of the zeroth-order Bessel function to find
\begin{align}
\mathcal{F}_2 &= e^{- 2\, |\alpha|^2} \biggl[ e^{2 \, |\alpha|^2\, e^{- \pi \tilde{\kappa}_{\rm{c}}}}  - I_0 (2 \, |\alpha|^2 \, e^{- \pi \tilde{\kappa}_{\rm{c}}})  \nonumber \\
&\qquad \qquad + \sum_{q = 1}^\infty \frac{( \tilde{\kappa}_{\rm{c}} |\alpha|^2)^q}{q!} \int^{2\pi}_0 \mathrm{d}\tau^{(1)} \cdots \mathrm{d}\tau^{(q)} e^{- \tilde{\kappa}_{\rm{c}} \sum_{p = 1}^q \tau^{(q)} } \left( e^{2 \, |\alpha|  e^{- \pi \tilde{\kappa}_{\rm{c}}}  \cos\left( 2  \sum_{p = 1}^q A( \tau^{(q)} ) \right) } - I _0 ( 2 \, |\alpha|e^{- \pi \tilde{\kappa}_{\rm{c}}}) \right) \biggr]. 
\end{align}
Rearranging, this expression can be written as 
\begin{align}
\mathcal{F}_2 &=  e^{- 2\, |\alpha|^2} \, e^{2 \, |\alpha|^2 e^{- \pi \tilde{\kappa}_{\rm{c}}}} \biggl[ 1 + \sum_{q = 1}^\infty \frac{( \tilde{\kappa}_{\rm{c}} |\alpha|^2)^q}{q!} e^{2 \, |\alpha|^2 e^{- \pi \tilde{\kappa}_{\rm{c}}} \left[ \cos\left( 2 \sum_{p = 1}^q A( \tau^{(q)} ) \right) - 1\right]}\biggr] \nonumber \\
&-e^{- 2\, |\alpha|^2} \left[ 1 +  \sum_{q = 1}^\infty \frac{( \tilde{\kappa}_{\rm{c}} |\alpha|^2)^q}{q!}
\int^{2\pi}_0 \mathrm{d}\tau^{(1)} \cdots \mathrm{d}\tau^{(q)} e^{- \tilde{\kappa}_{\rm{c}} \sum_{p = 1}^q \tau^{(q)} } \right] I_0 (2 \, |\alpha|^2 e^{- \pi \tilde{\kappa}_{\rm{c}}}) .
\end{align}
The integral in the second term of $\mathcal{F}_2$ can now be evaluated using~\eqref{app:eq:evaluated:integral} to find 
\begin{align} \label{app:eq:intermediate:terms}
\mathcal{F}_2 &=  e^{- 2\, |\alpha|^2} \, e^{2 \, |\alpha|^2 e^{- \pi \tilde{\kappa}_{\rm{c}}}} \biggl[ 1 + \sum_{q = 1}^\infty \frac{( \tilde{\kappa}_{\rm{c}} |\alpha|^2)^q}{q!} e^{2 \, |\alpha|^2 e^{- \pi \tilde{\kappa}_{\rm{c}}} \left[ \cos\left( 2 \sum_{p = 1}^q A( \tau^{(q)} ) \right) - 1\right]}\biggr] \nonumber \\
&\quad - e^{- 2\, |\alpha|^2} \left[ 1 +  \sum_{q = 1}^\infty \frac{ |\alpha|^{2q}}{q!}
 \left( 1 - e^{- 2\pi \tilde{\kappa}_{\rm{c}}} \right)^q \right] I_0 (2 \, |\alpha|^2 e^{- \pi \tilde{\kappa}_{\rm{c}}})  .
\end{align}
The sum over $q$ in the second term can then be evaluated to
\begin{align}
1 + \sum_{q = 1}^\infty \frac{|\alpha|^{2q}}{q!} \left( 1 - e^{ - 2\pi \tilde{\kappa}_{\rm{c}}} \right)^q = e^{|\alpha|^2 \left(1-e^{-2 \pi  \tilde{\kappa}_{\rm{c}} }\right)}.
\end{align}
With this, we see that the last term of $\mathcal{F}_2$ cancels the first term in~\eqref{app:eq:fidelity:Bessel:functions}, and we are finally left with 
\begin{align}  \label{app:eq:simplified:fidelity}
\mathcal{F}(2\,\pi) = e^{- 2 |\alpha|^2 ( 1 - e^{- \pi \tilde{\kappa}_{\rm{c}}})} \left[ 1 + \sum_{q = 1}^\infty \frac{(\tilde{\kappa}_{\rm{c}} |\alpha|^2)^q}{q!} \int^{2\pi}_0 \mathrm{d}\tau^{(1)} \cdots \tau^{(q)} e^{- \tilde{\kappa}_{\rm{c}} \sum_{p = 1}^q \tau^{(p)} } \, e^{- 4 \,  |\alpha|^2 e^{- \pi \tilde{\kappa}_{\rm{c}}} \sin^2 \left( \sum_{p = 1}^q A(\tau^{(p)} ) \right)} \right], 
\end{align}
where we have used the double-angle formula $\cos(2x ) -1= -2\,\sin^2x $.

This expression makes evident the orders of $\tilde{\kappa}_{\rm{c}}$ and of $\tilde{g}_0$ that enter into the expression [note that $A(\tau') \propto \tilde{g}_0^2$, which can be determined from inspection of the integrals in~\eqref{eq:definition:of:F:coefficients}]. Therefore, it lends itself well to perturbative expansions for small $\tilde{\kappa}_{\rm{c}} |\alpha|^2$ and weakly coupled systems where $\tilde{g}_0 \ll 1$.

\subsection{Bounding the fidelity} \label{app:bounding:fidelity}
The expression in~\eqref{app:eq:simplified:fidelity} can be bounded from above and below without assuming specific values of $\tilde{g}_0$, $\tilde{\kappa}_{\rm{c}}$, or $\alpha$. We first note from~\eqref{app:eq:simplified:fidelity} that the second exponential contains an argument of the form $-\sin^2\left[ A\left( \tau \right) \right]$ in an exponential, and therefore the integral itself is maximized when $A(\tau) = 0$ and minimizsed when $A(\tau) = \pi/2$. Just like in Appendix~\ref{app:quadratures}, we use the fact that the integral is bounded by the maximum value of its argument:
\begin{align}
\int \mathrm{d}x f_{\rm{min}}(x) \leq \int \mathrm{d}x \, f(x) \leq \int \mathrm{d}x \, f_{\rm{max}}(x). 
\end{align}
Considering the upper bound, where we set $A(\tau) = 0$, we have
\begin{align}
\mathcal{F} \leq e^{- 2 |\alpha|^2 ( 1 - e^{- \pi \tilde{\kappa}_{\rm{c}}})} \left[ 1 + \sum_{q = 1}^\infty \frac{(\tilde{\kappa}_{\rm{c}} \, |\alpha|^2)^q}{q!} \int^{2\pi}_0 \mathrm{d}\tau^{(1)} \ldots \tau^{(q)} e^{- \tilde{\kappa}_{\rm{c}} \sum_{p = 1}^q \tau^{(p)} }  \right] = e^{- |\alpha|^2 (1 - e^{- \pi \tilde{\kappa}_{\rm{c}}})^2}.
\end{align}
Note again that this expression is valid for any parameter regime, yet gives us a simple intuitive notion of the allowed values of $\tilde{\kappa}_{\rm{c}}$ given a specific $\alpha$ and desired fidelity. 

For the lower bound, we instead look at the minimum value of the integral, when $\sin^2[ A(\tau)] = 1$. We find
\begin{align}
\mathcal{F} &\leq e^{- 2 |\alpha|^2 ( 1 - e^{- \pi \tilde{\kappa}_{\rm{c}}})} \left[ 1 + \sum_{q = 1}^\infty \frac{(\tilde{\kappa}_{\rm{c}} |\alpha|^2)^q}{q!} \int^{2\pi}_0 \mathrm{d}\tau^{(1)} \ldots \tau^{(q)} e^{- \tilde{\kappa}_{\rm{c}} \sum_{p = 1}^q \tau^{(p)} } \, e^{- 4 |\alpha|^2 e^{- \pi \tilde{\kappa}_{\rm{c}}} } \right] \nonumber \\
&= e^{- |\alpha|^2 ( 1 - e^{- \pi \tilde{\kappa}_{\rm{c}}} )} \left[ 1 + e^{-4 |\alpha|^{2} e^{- \pi \tilde{\kappa}_{\rm{c}}}} \sum_{q = 1}^\infty \frac{ |\alpha|^{2q}}{q!} \left( 1 - e^{ - 2 \pi \tilde{\kappa}_{\rm{c}}} \right)^q \right]. 
\end{align}
This sum evaluates to 
\begin{align}
\sum_{q = 1}^\infty \frac{ |\alpha|^{2q}}{q!} \left( 1 - e^{ - 2 \pi \tilde{\kappa}_{\rm{c}}} \right)^q  = e^{| \alpha | ^2\left(1-e^{-2 \pi  \tilde{\kappa}_{\rm{c}} }\right) }-1, 
\end{align}
which allows us to write the fidelity as 
\begin{align}
\mathcal{F} &\leq e^{- |\alpha|^2 ( 1 - e^{- \pi \tilde{\kappa}_{\rm{c}}} )} \left[ 1 + e^{-4 |\alpha|^{2} e^{- \pi \tilde{\kappa}_{\rm{c}}}} \left( e^{| \alpha |^2\left(1-e^{-2 \pi  \tilde{\kappa}_{\rm{c}} }\right) }-1 \right) \right] \nonumber \\
&= 2 \, e^{-2 \,  |\alpha|^2} \, \sinh(2 \, |\alpha|^2 e^{- \pi \tilde{\kappa}_{\rm{c}}} ) + e^{- |\alpha|^2 ( 1 - e^{- \pi \tilde{\kappa}_{\rm{c}}})}. 
\end{align}
Again, this bound is completely general for all values of $\tilde{g}_0$, $\tilde{\kappa}_{\rm{c}}$, and $\alpha$. 

Summarizing our results, we have shown that the fidelity can be upper and lower bounded as 
\begin{align} \label{app:eq:fidelity:bounds}
2 \, e^{- 2 \, |\alpha|^2} \,\sinh (2 \,|\alpha|^2 e^{- \pi \tilde{\kappa}_{\rm{c}}})  + e^{- |\alpha|^2 ( 1 + e^{- \pi \tilde{\kappa}_{\rm{c}}})^2 } \leq \mathcal{F} \leq e^{- |\alpha|^2 ( 1 - e^{- \pi \tilde{\kappa}_{\rm{c}}}) ^2 }.
\end{align}

\end{document}